\documentclass[review,pdflatex,sn-mathphys-num]{sn-jnl}

\usepackage{graphicx}%
\usepackage{multirow}%
\usepackage{float}
\usepackage{bm}
\usepackage{amsmath,amssymb,amsfonts}%
\usepackage{amsthm}%
\usepackage[title]{appendix}%
\usepackage{xcolor}%
\usepackage{textcomp}%
\usepackage{manyfoot}%
\usepackage{booktabs}%
\usepackage{algorithm}%
\usepackage{algorithmicx}%
\usepackage{algpseudocode}%
\usepackage{listings}%
\usepackage{siunitx}
\usepackage{mathtools}
\usepackage{hyperref}

\DeclareMathOperator{\Tr}{Tr}

\theoremstyle{thmstyleone}%

\theoremstyle{thmstyletwo}%
\newtheorem{remark}{Remark}%

\theoremstyle{thmstylethree}%

\begin{document}

\title[Fundamental limits of stability inference in high-dimensional complex systems]{Fundamental limits of stability inference in high-dimensional complex systems}

\author[1]{\fnm{Michela} \sur{Costa}}
\affil[1]{
\orgdiv{Department of Social and Economic Sciences},
\orgname{Sapienza University of Rome},
\country{Italy}
}

\author[2]{\fnm{Kentaro} \sur{Hoshisashi}}
\affil[2]{\orgdiv{Department of Computer Science}, \orgname{University College London}, \orgaddress{\city{London}, \postcode{WC1E 6BT}, \country{United Kingdom}}}

\author[3]{\fnm{Flaviano} \sur{Morone}}
\affil[3]{\orgdiv{Department of Physics}, \orgname{New York University}, \orgaddress{\city{New York}, \state{NY}, \country{USA}}}

\author[4]{\fnm{Tim} \sur{Rogers}}
\affil[4]{\orgname{University of Bath}, \country{United Kingdom}}

\author[2]{\fnm{Paolo} \sur{Barucca}}

\abstract{Many complex systems, including ecosystems, neural circuits, 
and financial markets, are inferred to operate close to a threshold of 
instability, at which a small perturbation can propagate across the 
entire system. This proximity is often interpreted as functionally 
advantageous, yet it poses a question common to all these fields: from 
a finite, noisy recording, how precisely can the distance of a system 
from that threshold be estimated?

Using the multivariate Ornstein--Uhlenbeck process as the canonical
linear model of relaxation near a stable fixed point, we show that the
attainable precision is governed by three factors: an effective
measurement budget, set by the number of samples relative to the system
dimension and the sampling interval; the signal-to-noise ratio, given
by the magnitude of deterministic interactions relative to stochastic
forcing; and the distance to criticality, which simultaneously
sets the system's correlation times and degrades both of the preceding
factors. As the slowest dynamical mode softens near the 
threshold, the curvature of the log-likelihood flattens along the 
direction that determines stability, so that the relative uncertainty 
on the estimated distance diverges as that distance vanishes. 
Critically, temporal correlations near instability reduce the effective 
number of independent observations far below the nominal sample count, 
and inference breaks down when this effective count falls below the 
system dimension, even when the raw data volume appears sufficient. 
A direct consequence is the existence of an optimal sampling interval 
that diverges as the system approaches criticality, with practical 
implications for experimental design.

We validate these predictions in Monte Carlo simulations for the
Gaussian Orthogonal Ensemble (GOE), modelling reversible processes
with symmetric interactions, and the Ginibre ensemble, covering
irreversible processes with asymmetric interactions. Applying the framework to three empirical datasets, we find that 
the distance to instability is precisely resolved for financial 
volatility data during the COVID-19 onset; for plankton abundance 
data stability of the 
system cannot be diagnosed; for intracranial EEG during epileptic 
seizures stability can be reliably diagnosed but the precise 
distance to instability cannot be resolved.}


\maketitle

\section{Introduction}\label{sec:intro}
A recurring proposition across the sciences is that complex systems
operate near the boundary of stability. Since May's analysis of whether a
large complex system can be stable~\cite{may1972will}, random matrix
arguments have related this boundary to the statistical structure of the
interactions in ecosystems~\cite{allesina2012stability} and in large
economies~\cite{moran2019may}. Ecosystems, neural circuits and financial
markets share a common set of signatures of
criticality~\cite{bialek2012biophysics,mora2011biological,scheffer2009early}:
long relaxation times, broad (scale-free) fluctuation distributions, and
heightened sensitivity to perturbations. Proximity to instability has been
argued to confer adaptability and responsiveness while also increasing the
susceptibility to system-wide cascades. This proximity need not arise from
a self-organized criticality scenario: it may result from evolutionary
fine-tuning, from the statistical structure of random interactions as in
May's argument~\cite{may1972will}, or from the system operating in a
parameter regime close to the boundary for other reasons. In any of these
cases, the quantity of interest is the \emph{distance} of the system from
the threshold, and the central question is whether this distance can be
estimated from the data the system generates.

We ask how precisely the distance to instability
can be inferred from a finite, noisy time series,
and show that the limit on that precision is not
a numerical artifact but a structural property of
criticality itself, consistent with the broader
observation that inferred models tend toward
criticality~\cite{mastromatteo2011criticality}.
The uncertainty on the estimated distance grows
as the system approaches the critical point.

To make the question precise while retaining generality, we adopt the
multivariate Ornstein--Uhlenbeck (OU) process, the minimal linear model of
stochastic relaxation toward a stable fixed point,
\begin{equation}
d\mathbf{X}(t) = -\mathbf{A}\,\mathbf{X}(t)\,dt + \boldsymbol{\eta}(t),
\label{eq:ou}
\end{equation}
where $\mathbf{X}(t)\in\mathbb{R}^N$ is the state vector, the drift matrix
$\mathbf{A}\in\mathbb{R}^{N\times N}$ encodes the interactions, and
$\boldsymbol{\eta}\in\mathbb{R}^{N}$ is Gaussian white noise such that $\langle \boldsymbol{\eta}(t)\,\boldsymbol{\eta}(t')^\top \rangle = 2\mathbf{B}\,\delta(t-t')$ where
$\mathbf{B}\in\mathbb{R}^{N\times N}$ is the diffusion matrix.

Stability is determined by the spectrum of $\mathbf{A}$: the process
has a stationary distribution if and only if all eigenvalues of
$\mathbf{A}$ have positive real part, and the system approaches
instability as $r := \min_i\Re[\lambda_i(\mathbf{A})] \to 0^+$.

In high-dimensional systems the precise structure of the interaction
matrix $\mathbf{A}$ is rarely accessible; what is typically known
are its statistical properties: the distribution of coupling
strengths, the degree of symmetry, and the spectral density.
Random matrix theory (RMT) provides a natural framework to
characterise the typical behaviour of such systems: rather than
fixing a single $\mathbf{A}$, one draws it from an ensemble and
asks what properties hold with high probability. This approach is
particularly useful near instability, where the relevant quantity, the smallest eigenvalue of $\mathbf{A}$, is a spectral edge statistic whose fluctuations are governed by universal laws
independent of the microscopic details of the couplings.

To place the system a controlled distance from instability, we draw
$\mathbf{A}$ from a random matrix ensemble tuned by a single control
parameter $c$, so that the ensemble-averaged distance to criticality,
\begin{equation}
r(c) := \mathbb{E}_{P(\mathbf{A})}\!\left[\min_{\alpha}
\Re(\lambda_\alpha(\mathbf{A}))\right]
\xrightarrow{\;c\to c_{\mathrm{crit}}\;} 0^+,
\label{eq:lmin_c}
\end{equation}
vanishes at a critical value $c_{\mathrm{crit}}$ whose specific form
depends on the matrix ensemble under consideration.

Three factors govern the attainable precision. The first is the
\emph{effective measurement budget}: not the number of samples alone,
but the number of samples relative to the system dimension $N$,
together with the sampling interval $\Delta t$. Sampling too coarsely
discards dynamical information, whereas sampling much faster than the
relaxation times yields strongly correlated, redundant observations.
The second is the \emph{signal-to-noise ratio}: the magnitude of the
deterministic interactions relative to the stochastic forcing
$\mathbf{B}$.The third, and the central quantity of interest in this work, 
is the \emph{distance to criticality} $r$ itself: as the slowest 
dynamical mode, the eigenvector of $\mathbf{A}$ associated 
with the smallest eigenvalue $r$, softens, two effects combine. The stationary fluctuations grow, obscuring the deterministic
signal, and the relaxation times diverge, so that consecutive
observations become nearly indistinguishable and the effective number
of independent samples collapses, making the first two factors
progressively worse as $r \to 0$. The three factors are therefore not
independent: proximity to instability simultaneously degrades the
signal-to-noise ratio and inflates the temporal correlations that
reduce the effective measurement budget. A further consideration is
that at criticality the linear, stationary description underlying
Eq.~\eqref{eq:ou} ceases to be valid, since relaxation times diverge
and the process approaches non-stationarity; the model used to define
the distance to instability is therefore least applicable in the
regime where that distance is smallest.

We show that these factors combine quantitatively. A quadratic
approximation of the log-likelihood reduces the estimation problem to
the curvature of the inference landscape, which admits a closed-form
expression and vanishes along the soft mode that defines criticality.
The relative uncertainty on the estimated distance to instability
diverges as that distance vanishes, and this divergence is not a
numerical artefact but a structural property of the inference problem:
the signal vanishes faster than the noise as the system approaches
the threshold.
A consequence is the existence of an effective
sample-to-dimension ratio that accounts for temporal correlations 
and is strictly smaller than the nominal one. This gives rise to 
three distinct thresholds of increasing stringency. The first is 
the breakdown threshold: when temporal correlations reduce the 
effective sample count below the system dimension, the empirical 
covariance matrix becomes singular and inference breaks down 
entirely, even when the nominal sample count appears sufficient. The second 
is the stability diagnosis threshold: above it, the sign of $r$, 
and hence whether the system is stable or unstable, cannot be 
reliably determined from the data. The third is the resolution 
threshold: below it, the distance to instability is 
quantitatively resolved with better than $10\%$ relative 
precision. Together these thresholds define two resolvability 
phase diagrams in the space of the coupling strength, the 
sample-to-dimension ratio, and the sampling interval, which 
reveal an optimal sampling interval that diverges as the system 
approaches criticality, with direct implications for experimental 
design in neuroscience, ecology, and finance.

We validate these predictions in Monte Carlo simulations for two 
random matrix ensembles: the Gaussian Orthogonal Ensemble (GOE), 
which models reversible processes with symmetric drift matrix 
$\mathbf{A}$, and the Ginibre ensemble, which covers irreversible, 
non-normal processes with asymmetric $\mathbf{A}$. Both ensembles 
are characterised theoretically, but closed-form analytical 
expressions for the resolvability boundary are derived only for 
the GOE case, where the symmetry of $\mathbf{A}$ renders the 
eigenvalue statistics tractable; for the Ginibre ensemble we 
provide numerical results. We then apply the framework to three 
empirical datasets, characterising each according to two 
complementary criteria: a \emph{stability diagnosis} criterion, 
which asks whether the sign of $r$, and hence whether the 
system is stable ($r>0$) or unstable ($r<0$), can be reliably 
determined from the data; this corresponds to the threshold 
$\sigma_r/r = 1$, at which the probability of falsely diagnosing 
an unstable system is $P_{\mathrm{false}} = \Phi(-1) \approx 16\%$; 
and a \emph{resolution} criterion, which asks whether the magnitude 
of $r$ can be estimated with better than $10\%$ relative precision 
($\sigma_r/r \leq 0.1$).
Financial volatility data from a panel 
of large-cap US equities during the onset of the COVID-19 pandemic 
meets both criteria ($\sigma_r/r = 0.04$): the system is 
unambiguously diagnosed as stable and its distance to instability 
is quantitatively resolved. Plankton abundance data meets neither 
criterion ($\sigma_r/r = 1.56$): temporal correlations reduce the 
effective sample count below the system dimension despite a 
nominally adequate data volume, so that even the stability of the 
system cannot be reliably diagnosed. Intracranial EEG data during 
epileptic seizures meets the diagnosis criterion but not the 
resolution criterion ($\sigma_r/r = 0.37$): the system is 
reliably diagnosed as stable, but the precise distance to 
instability is not quantitatively resolved.

\section{Results}\label{sec:results}
Suppose that a trajectory of the multivariate Ornstein--Uhlenbeck 
process \eqref{eq:ou} is observed at discrete times, yielding a 
dataset of $M$ observations in an $N$-dimensional state space. 
For a sampling interval $\Delta t$, the process admits the 
discrete-time representation
\begin{equation}
\mathbf{X}_{k+1}
=
\mathbf{Q}\,\mathbf{X}_k
+
\boldsymbol{\varepsilon}_k,
\end{equation}
with
\begin{equation}
\mathbf{Q}=e^{-\mathbf{A}\Delta t},
\end{equation}
and Gaussian innovations
\begin{equation}
\boldsymbol{\varepsilon}_k
\sim
\mathcal{N}\!\left(0,\,\boldsymbol{\Sigma}_{\Delta t}\right),
\qquad
\boldsymbol{\Sigma}_{\Delta t}
=
\boldsymbol{\Sigma}_{\infty}
-
\mathbf{Q}\boldsymbol{\Sigma}_{\infty}\mathbf{Q}^\top.
\end{equation}
The discretely sampled process is therefore fully characterized by 
the parameter set $\boldsymbol{\theta}=(\mathbf{Q},
\boldsymbol{\Sigma}_{\Delta t})$, which is directly accessible from 
the observed data.
Once these quantities have been estimated, the continuous-time 
parameters $(\mathbf{A}, \mathbf{B})$ can be recovered.

Throughout this work we consider uniform priors, so that inference 
is entirely determined by the likelihood of the observed trajectory 
(the derivation of the likelihood, estimators and their sufficient 
statistics is reported in Methods). The posterior distribution is 
maximized by the maximum a posteriori (MAP) estimator 
$\hat{\boldsymbol{\theta}}=(\hat{\mathbf{Q}},
\hat{\boldsymbol{\Sigma}}_{\Delta t})$.

As a measure of the accuracy of our estimates, we use the Fisher 
information matrix $\mathcal{I} = 
\mathcal{I}(\boldsymbol{\Sigma}_{\Delta t},
\boldsymbol{\Sigma}_\infty)$, which sets 
the Cram\'er--Rao lower bound on the variance of any unbiased 
estimator. It admits a block-diagonal 
structure separating the contributions of $\mathbf{Q}$ and 
$\boldsymbol{\Sigma}_{\Delta t}$,
\begin{equation}
\mathcal{I} =
\begin{pmatrix}
\mathcal{I}^{(\Sigma\Sigma)} & \mathbf{0} \\
\mathbf{0} & \mathcal{I}^{(QQ)}
\end{pmatrix}
\in \mathbb{R}^{2N^2 \times 2N^2},
\qquad
\mathcal{I}^{(\Sigma\Sigma)} = \boldsymbol{\Sigma}_{\Delta t} 
\otimes \boldsymbol{\Sigma}_{\Delta t},
\qquad
\mathcal{I}^{(QQ)} = 2\,\mathbf{J} \otimes \boldsymbol{\Sigma}_\infty,
\label{eq:fisher_blocks}
\end{equation}
where $\mathbf{J} = \boldsymbol{\Sigma}_{\Delta t}^{-1}$.
Since the true parameters are unknown in practice, we also consider 
the plug-in estimator $\hat{\mathcal{I}} = 
\mathcal{I}(\hat{\boldsymbol{\Sigma}}_{\Delta t},
\hat{\boldsymbol{\Sigma}}_\infty)$, obtained by replacing the 
population covariances with their empirical counterparts.

\begin{figure}[t]
\centering
\includegraphics[width=0.7\textwidth]{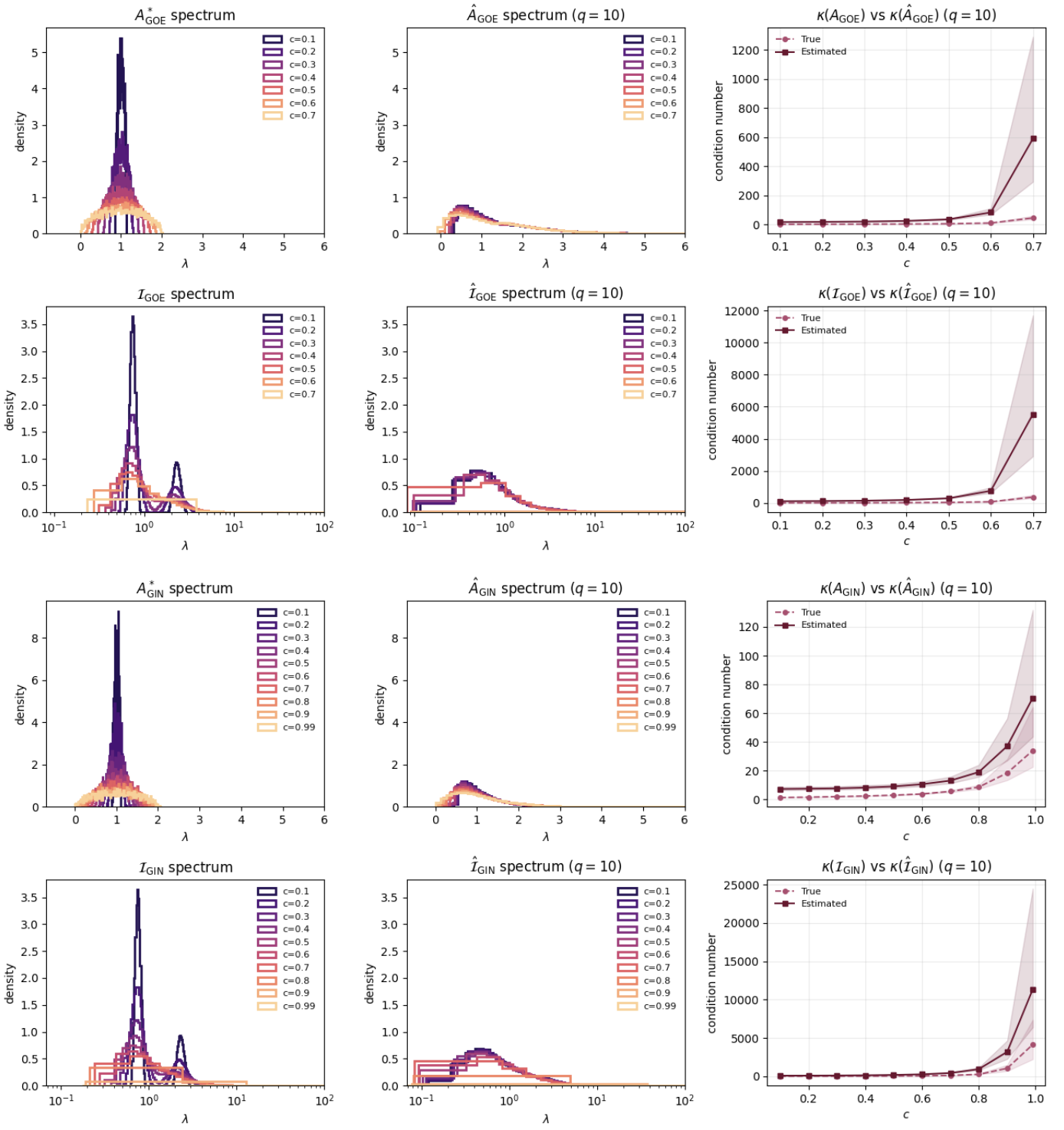}
\caption{
Eigenvalue spectra and condition numbers for the GOE (rows 1--2)
and Ginibre (rows 3--4) ensembles, each with $N=100$, $q=10$,
$\Delta t=1$, $n_A=30$ independent realisations, and $n_{\rm sim}=30$
trajectories per realisation. For each ensemble, the left column
shows the true spectrum, the center column the estimated spectrum,
and the right column the condition number as a function of $c$.
\textbf{Row 1 (GOE, drift matrix):} true spectrum of $\mathbf{A}$,
plug-in estimator $\hat{\mathbf{A}}$, and condition numbers
$\kappa(\mathbf{A})$ and $\kappa(\hat{\mathbf{A}})$.
\textbf{Row 2 (GOE, Fisher information):} true Fisher matrix
$\mathcal{I}$, plug-in estimator $\hat{\mathcal{I}}$, and their
condition numbers.
\textbf{Rows 3--4} repeat the same comparisons for the Ginibre
ensemble; spectra show the real part of the eigenvalues.
Colors denote coupling strengths $c\in\{0.1,0.2,\ldots,0.7\}$
for the GOE and $c\in\{0.1,0.2,\ldots,0.99\}$ for the Ginibre.
As $c$ increases toward $c_{\rm crit}$, the spectra broaden,
the Fisher matrix becomes increasingly anisotropic, and the
condition number grows rapidly in both ensembles. The Ginibre
ensemble reaches much larger condition numbers before breakdown
owing to its higher $c_{\rm crit} = 1$. Only stable
reconstructions satisfying $\lambda_{\min}(\hat{\mathbf{A}}) > 0$
are included.
}
\label{fig:spectra}
\end{figure}

As the system approaches the stability threshold, the smallest 
eigenvalue of $\mathbf{A}$ approaches zero and the associated 
relaxation time diverges. The corresponding dynamical mode becomes 
increasingly difficult to constrain statistically, generating a 
progressively flatter direction in the posterior landscape. 
Consequently, the spectrum of $\mathcal{I}$ broadens in both 
directions: the smallest eigenvalues decrease while the largest 
increase, so that the condition number 
\begin{equation}
    \kappa(\mathcal{I}) = \frac{\mu_{\max}(\mathcal{I})}{\mu_{\min}(\mathcal{I})}
\end{equation}
grows rapidly, where $\mu_{\max}(\mathcal{I})$ and $\mu_{\min}(\mathcal{I})$ 
are the largest and smallest eigenvalues of $\mathcal{I}$, 
respectively.
The posterior becomes increasingly anisotropic, with some 
directions remaining well-constrained while others become 
progressively undetermined.
Figure~\ref{fig:spectra} compares the spectra of 
$\mathbf{A}$ and $\hat{\mathbf{A}}$ with those of $\mathcal{I}$ 
and $\hat{\mathcal{I}}$. The estimated spectra are systematically 
broader than their theoretical counterparts owing to finite-sample 
fluctuations and the reduction of the effective number of 
independent observations caused by critical slowing down. As 
criticality is approached, these effects become increasingly 
pronounced, leading to
a loss of inferential precision.\\

The reduction of inferential precision near criticality has a 
precise quantitative origin beyond the growth of stationary 
fluctuations. The observed trajectory consists of $M$ snapshots 
separated by $\Delta t$, but these are not independent: the 
eigenvalues $\lambda_\alpha$ govern the decay rates of temporal 
correlations along each mode, and as the system approaches 
instability they become vanishingly small, making consecutive 
observations nearly indistinguishable and reducing the effective 
number of independent samples.

When $\mathbf{A}$ is symmetric, the OU process decouples into $N$
independent AR(1) processes, one per eigenvector. The effective
number of independent samples along mode $\alpha$ is
\begin{equation}
    M_{\rm eff}^\alpha
    =
    \frac{M}{\tau_\alpha},
    \qquad
    \tau_\alpha
    =
    \frac{1+e^{-2\lambda_\alpha \Delta t}}
         {1-e^{-2\lambda_\alpha \Delta t}},
    \label{eq:Meff_alpha}
\end{equation}
which vanishes as $M_{\rm eff}^\alpha \approx M\lambda_\alpha\Delta t$
in the slow-mode limit $\lambda_\alpha\Delta t \ll 1$. Since
$\lambda_{\min}(\mathbf{A})$ is extracted from the full estimated
matrix $\hat{\mathbf{A}}$, the relevant effective sample size is
a single global quantity giving more weight to the slow modes,
\begin{equation}
    M_{\rm eff}
    \approx
    \frac{M}{\langle\tau\rangle_{\rm w}},
    \qquad
    \langle\tau\rangle_{\rm w}(c, \Delta t)
    =
    \frac{\sum_{\alpha=1}^N \tau(\lambda_\alpha)^2}
         {\sum_{\alpha=1}^N \tau(\lambda_\alpha)},
    \label{eq:Meff_tau}
\end{equation}
so that $\langle\tau\rangle_{\rm w}$ 
depends on $c$ through the eigenvalues of $\mathbf{A}$, which 
is sampled from an ensemble with coupling strength $c$, and 
on $\Delta t$.

This has a powerful consequence: the original dataset of $M$ 
temporally and spatially correlated observations can be treated 
as an equivalent dataset of $M_{\rm eff}$ i.i.d.\ samples with 
spatial correlations encoded in $\Sigma_\infty$, placing the 
residual inference problem within the framework of deformed 
Wishart random matrix theory \cite{marchenko1967distribution}, with effective ratio 
$q_{\rm eff} = M_{\rm eff}/N$.\\

The ratio $q_{\rm eff} = M_{\rm eff}/N$ plays the role of an
effective sample-to-dimension ratio and determines the feasibility
of inference. When $q_{\rm eff} \leq 1$ the empirical covariance
matrix $\hat{T}_3$ is no longer invertible: its spectrum develops
vanishing and eventually negative eigenvalues, violating positive
definiteness. As a consequence the estimator $\hat{\mathbf{Q}} =
T_2 T_3^{-1}$ becomes ill-defined, the inferred drift matrix
$\hat{\mathbf{A}}$ develops eigenvalues crossing zero, and the
reconstructed system is falsely diagnosed as unstable. This
threshold $q_{\rm eff} = 1$ is therefore sharper than the naive
requirement $q = M/N > 1$: a near-critical system can enter the
regime $q_{\rm eff} < 1 < q$, where the nominal sample size appears
sufficient but temporal correlations render the observations
effectively redundant.\\

A central question is how well one can resolve the distance to
instability from a finite time series. Denoting by $r =
\lambda_{\min}(\boldsymbol{A})$ the smallest eigenvalue of the drift
matrix, the variance of its estimator $\hat{r}$ is
\begin{equation}
    \mathrm{Var}(\hat{r})
    = \kappa_W\,
    \frac{e^{2r\Delta t}-1}{M\,\Delta t^2},
    \label{eq:var_lambda}
\end{equation}
where $\kappa_W = (M-1)/(M-N-2)$ is the inverse-Wishart inflation
factor. When temporal correlations are present, $\kappa_W$ is
replaced by a larger effective factor. A naive compound-Wishart
approximation is obtained by replacing the nominal sample size
$M-1$ with the effective sample size $M_{\rm eff}$ from
\eqref{eq:Meff_tau},
\begin{equation}
    \kappa_{\rm cW}
    \approx \frac{M_{\rm eff} - 1}{M_{\rm eff} - N - 2},
    \label{eq:kcW_results}
\end{equation}
defined for $M_{\rm eff} > N+2$ and diverging as $M_{\rm eff}
\to (N+2)^+$.\\

In the slow-mode limit $r\Delta t \ll 1$, the relative uncertainty 
takes the form
\begin{equation}
    \frac{\sigma_r}{r}
    \approx
    \sqrt{\frac{2\kappa_{\rm cW}}{r\,M\Delta t}},
    \label{eq:divergence}
\end{equation}
which diverges as $r\to 0$. This follows from a fundamental 
asymmetry: the absolute fluctuations $\sigma_r\propto\sqrt{r}$ 
vanish more slowly than $r$ itself, so the signal-to-noise ratio 
$r/\sigma_r$ collapses. 
In the simplest case $\kappa_{\rm cW} \to \kappa_W$, the divergence 
goes as $r^{-1/2}$; when the full compound-Wishart correction is 
included, $\kappa_{\rm cW}$ itself grows as $r \to 0$, making the 
divergence faster than $r^{-1/2}$.

\begin{figure}[t]
\centering
\includegraphics[width=0.75\textwidth]{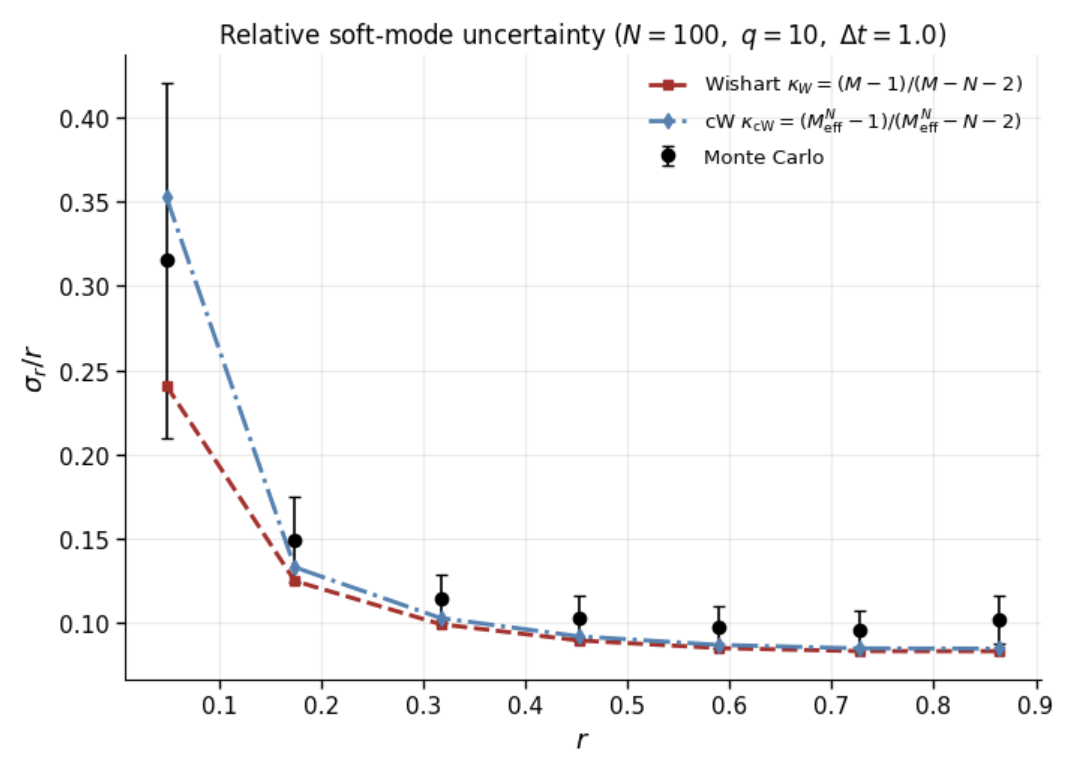}
\caption{
Relative uncertainty of the soft mode as a function of the 
distance to instability $r=\lambda_{\min}(A)$. The relaxation 
rate is estimated by projecting the inferred propagator $\hat{Q}$ 
onto the true soft eigenvector $u_*$. Black circles show Monte 
Carlo results; the red dashed curve corresponds to the Wishart 
prediction $\kappa_W$ and the blue dash-dotted curve to the 
compound-Wishart approximation $\kappa_{\rm cW}$. The 
compound-Wishart correction provides a better approximation 
to the empirical uncertainty than the naive Wishart bound, 
consistently capturing the growth of $\sigma_r/r$ as criticality 
is approached. The projection isolates the soft mode and removes 
the order-statistic fluctuations associated with 
$\lambda_{\min}(\hat{A})$.
}
\label{fig:projected_soft_mode_uncertainty}
\end{figure}

Figure~\ref{fig:projected_soft_mode_uncertainty} shows Monte Carlo
estimates of $\sigma_r/r$ alongside the two theoretical predictions:
the Wishart bound $\kappa_W$ and the compound-Wishart correction
$\kappa_{\rm cW}$. The simulations are in good agreement with
$\kappa_{\rm cW}$, confirming that the compound-Wishart correction
captures the dominant effect of temporal correlations. Concretely,
as the system approaches criticality $\sigma_r/r$ grows by an order
of magnitude; a system with $\sigma_r/r \sim 0.3$ carries a $30\%$
relative uncertainty on $\hat{r}$, meaning that the inferred
distance to instability is poorly resolved.

These results admit a direct operational interpretation. To draw
the resolvability boundary at finite $N$, we need the expected
distance to instability $\bar{r}(c,N)$, which receives a finite-size
correction from Tracy--Widom edge statistics,
\begin{equation}
    \bar{r}(c,N)
    =
    (1-\sqrt{2}\,c)
    -
    (\sqrt{2}\,c)^{1/3}\,\mu_{\mathrm{TW}}\,N^{-2/3},
    \qquad
    \mu_{\mathrm{TW}}\approx -1.21.
    \label{eq:TW_floor}
\end{equation}
Since $\mu_{\mathrm{TW}}<0$, finite systems are on average slightly
more stable than predicted by the large-$N$ relation
$r = 1-\sqrt{2}\,c$. Substituting $\bar{r}(c,N)$ into \eqref{eq:divergence} expresses 
$\sigma_r/r$ as a function of $(c, q, \Delta t)$; fixing this 
ratio to a prescribed threshold defines a resolvability boundary 
in the $(c, q, \Delta t)$ parameter space, which takes different 
forms for the diagnosis threshold ($\sigma_r/r = 1$) and the 
resolution threshold ($\sigma_r/r = 0.1$).

\begin{figure}[t]
\centering
\includegraphics[width=\textwidth]{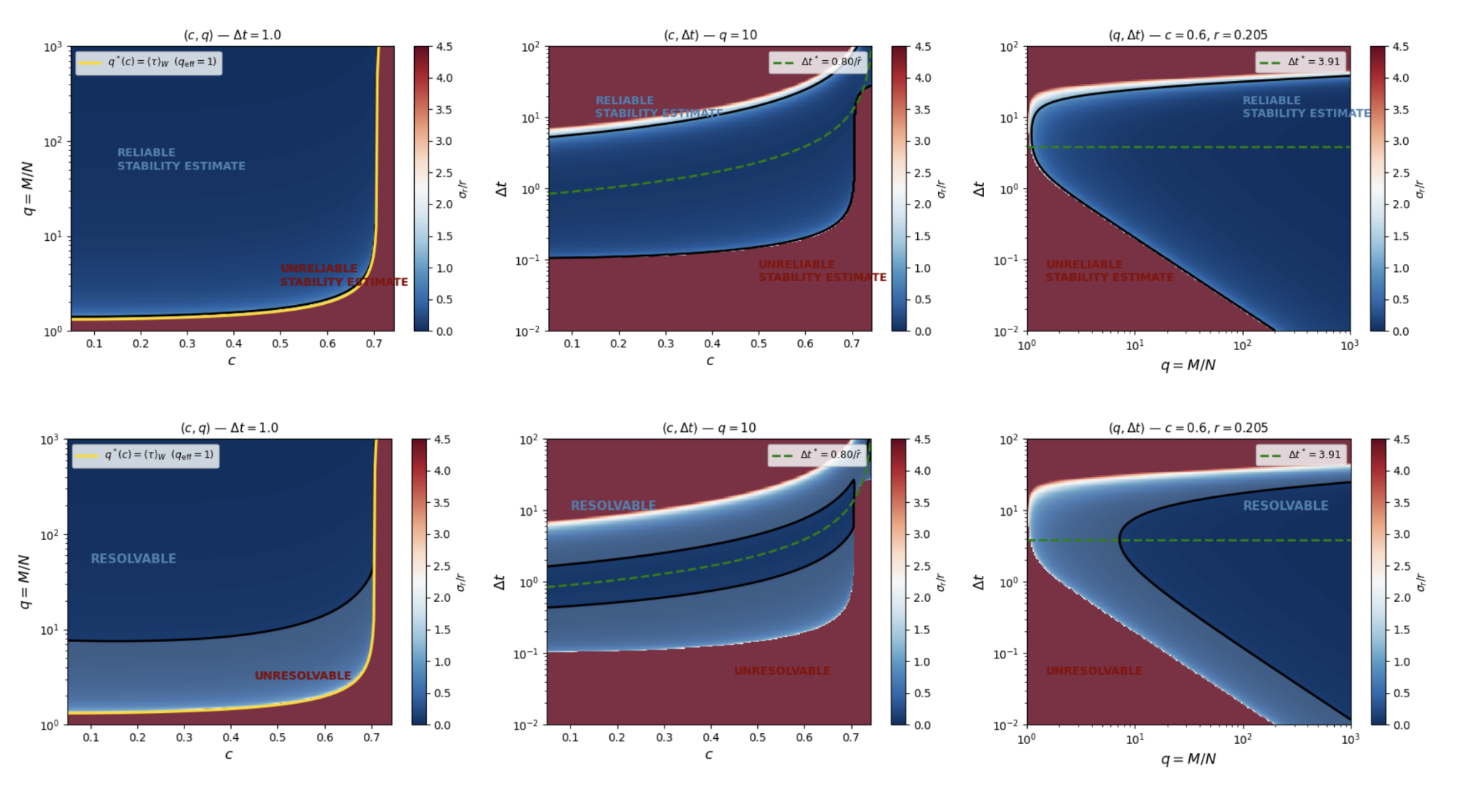}
\caption{
Stability inference and resolution phase diagrams for the GOE
ensemble ($N=100$, $\kappa_{\mathrm{cW}}$ inflation factor,
Tracy--Widom finite-size correction). Each panel shows the relative
uncertainty $\sigma_r/r$ as a function of two system or experimental
parameters, with the remaining parameter fixed.
\textbf{Top row (stability diagnosis boundary,} $\sigma_r/r = 1$\textbf{):}
regions above the boundary (red overlay) correspond to a
false-instability probability $P_{\mathrm{false}} \geq \Phi(-1)
\approx 16\%$, i.e.\ the data are insufficient to reliably
determine the sign of $r$.
\textbf{Bottom row (resolution boundary,} $\sigma_r/r = 0.1$\textbf{):}
regions above the boundary indicate that $r$ cannot be estimated
with better than $10\%$ relative precision; below it, the distance
to instability is quantitatively resolved.
In both rows, regions above the boundary correspond to the 
unresolvable regime:
\emph{left}, $(c,q)$ plane at fixed $\Delta t = 1$; the gold curve
marks $q^*(c) = \langle\tau\rangle_{\rm w}$, below which 
$q_{\mathrm{eff}} < 1$ and inference breaks down entirely, a 
stricter feasibility threshold than the naive requirement $q > 1$;
\emph{centre}, $(c,\Delta t)$ plane at fixed $q = 10$; the green dashed
line marks the optimal sampling interval $\Delta t^* \approx
0.80/\bar{r}$, which minimises $\sigma_r/r$ at fixed number of
samples $M$ and diverges as $r \to 0$, reflecting the fact that
near-critical systems require increasingly slow sampling to extract
independent information;
\emph{right}, $(q,\Delta t)$ plane at fixed $c = 0.6$; the green
dashed line again marks $\Delta t^*$, showing that the measurement
budget $q$ and the sampling rate must be jointly optimised to
achieve reliable inference.
}
\label{fig:phase_diagram}
\end{figure}

Three thresholds of increasing stringency characterise the
feasibility and quality of inference, and are all visible in
Figure~\ref{fig:phase_diagram}.
Throughout, we quantify the risk of misdiagnosis via the 
false-instability probability. Treating the estimator $\hat{r}$ 
as approximately Gaussian with mean $r$ and standard deviation 
$\sigma_r$, the probability of falsely declaring a stable system 
unstable is
\begin{equation}
    P_{\mathrm{false}}
    = \Phi\!\left(-\frac{r}{\sigma_r}\right),
\end{equation}
where $\Phi$ is the standard normal CDF.\\
The most fundamental is the limit $q_{\mathrm{eff}} = 1$, marked
by the gold curve in the left panels, corresponding to
\begin{equation}
    q_{\rm eff} = \frac{q}{\langle\tau\rangle_{\rm w}} = 1,
\end{equation}
where $\langle\tau\rangle_{\rm w}$ is uniquely determined by $c$ 
and $\Delta t$. 
The threshold value $q^*(c) = \langle\tau\rangle_{\rm w}$ therefore 
depends only on $c$. At this threshold $\sigma_r/r \to \infty$, so that
$P_{\mathrm{false}} = \Phi(-r/\sigma_r) \to \Phi(0) = 50\%$:
the estimated distance to instability is equally likely to be
positive or negative, regardless of the true value of $r$, and
inference is impossible. A
near-critical system can satisfy the nominal requirement $q > 1$
yet have $q_{\mathrm{eff}} < 1$, because temporal correlations
render the observations effectively redundant and drive the
effective sample count below the system dimension.

A less stringent but practically important threshold is
$\sigma_r/r = 1$, shown in the top row of
Figure~\ref{fig:phase_diagram}. At $\sigma_r/r = 1$ this
gives $P_{\mathrm{false}} = \Phi(-1) \approx 16\%$: more than one
in six estimates incorrectly diagnoses instability. This is a
conventional threshold whose exact value depends on the Gaussian
approximation; what is unambiguous is that it lies strictly between
$0\%$ and $50\%$, i.e.\ strictly inside the feasible region
$q_{\mathrm{eff}} > 1$. It is also important to note that
$\sigma_r/r = 1$ signals a loss of \emph{precision} on $r$, not
a loss of the ability to detect instability altogether: resolving
the exact value of $r$ is a harder problem than merely determining
its sign, and it is the former that breaks down first as the system
approaches criticality.

The strictest criterion shown is $\sigma_r/r = 0.1$, displayed
in the bottom row of Figure~\ref{fig:phase_diagram}. Below this
boundary $r$ can be estimated with better than $10\%$ relative
precision, providing a quantitative resolution threshold that is
independent of any probabilistic approximation.

In practice, $r$ is unknown and must be replaced by $\hat{r}$
to evaluate these criteria. If $\hat{r}$ places the system in
the unresolvable region this serves as a warning; if it places
the system in the resolvable region the estimate is
self-consistently trustworthy.

The $(c,\Delta t)$ plane (centre panels) reveals an optimal
sampling interval
\begin{equation}
    \Delta t^* \approx \frac{0.80}{\bar{r}},
    \label{eq:dt_opt}
\end{equation}
which diverges as $r\to 0$: near-critical systems require
increasingly slow sampling to extract independent information,
incompatible with a fixed observation window. The $(q,\Delta t)$
plane (right panels) confirms that the measurement budget and
sampling rate must be jointly optimised.

\begin{figure}[t]
\centering
\includegraphics[width=\textwidth]{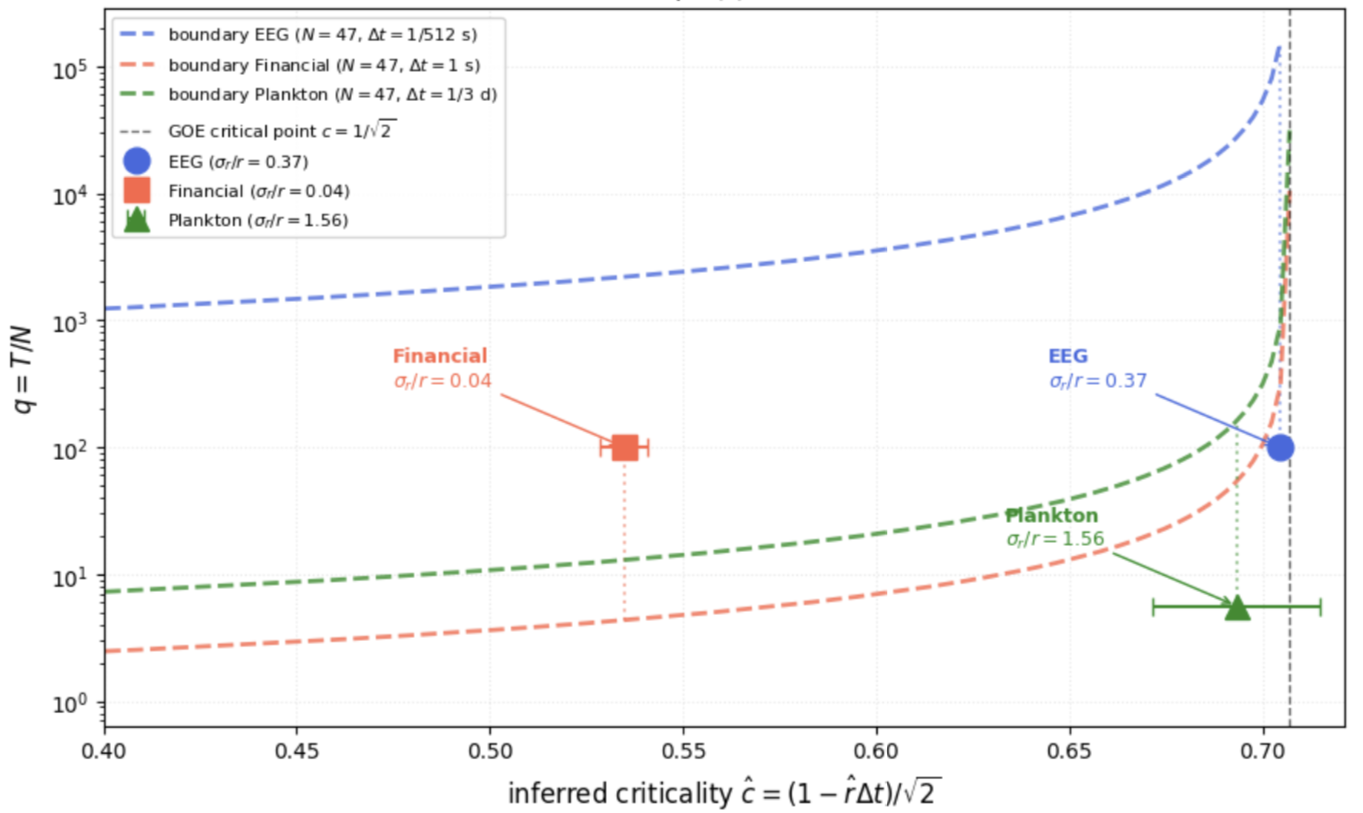}
\caption{
Resolvability phase diagram for three empirical datasets, overlaid
on the GOE boundary. Each dashed curve is the dataset-specific
resolvability boundary $q^*(c)$ defined by $\sigma_r/r = 1$,
computed using the system-specific parameters $(N, \Delta t)$;
points above the curve are resolvable. The financial volatility
series (red, $N=47$, $\Delta t = 1\,$s) lies well above its
boundary with $\sigma_r/r = 0.04$, indicating precise resolution
of the distance to instability. The intracranial EEG (blue,
$N=47$, $\Delta t = 1/512\,$s) sits close to its boundary with
$\sigma_r/r = 0.37$: the stability can be diagnosed but the
precise value of $r$ is poorly resolved. The plankton series
(green, $N=47$, $\Delta t = 1/3\,$d) falls below its boundary
with $\sigma_r/r = 1.56$, placing it in the unresolvable regime
where temporal correlations render the observations effectively
redundant. Horizontal error bars denote $\pm\sigma_r$ on the
inferred criticality $\hat{c} = (1 - \hat{r}\Delta t)/\sqrt{2}$.
The vertical dashed line marks the GOE critical point
$c = 1/\sqrt{2}$.
}
\label{fig:phase_diagram_datasets}
\end{figure}

\textbf{Application to real datasets.}
To test the framework on empirical data, we apply our inference
procedure to three multivariate time series from qualitatively
distinct domains. Each dataset is mapped onto the $(\hat{c}, q)$
plane relative to its own resolvability boundary, computed at
the dataset-specific $(N, \Delta t)$
(Figure~\ref{fig:phase_diagram_datasets}).

\paragraph{Financial volatility}
The financial volatility panel ($N=47$, $\Delta t = 1\,$s,
see Methods) sits clearly above the resolvability boundary,
with $\sigma_r/r = 0.04$: the distance to instability is
resolved with only $4\%$ relative uncertainty, confirming
reliable and precise inference despite the high-dimensional,
rapidly fluctuating nature of the system.

\paragraph{Plankton abundance}
The plankton dataset ($N=47$, $\Delta t = 1/3\,$d, see Methods)
falls in the unresolvable region, with $\sigma_r/r = 1.56$:
the uncertainty exceeds the signal itself, and the inference
is unreliable.

\paragraph{EEG during epileptic seizures}
The intracranial EEG dataset ($N=47$, $\Delta t = 1/512\,$s,
see Methods) gives $\sigma_r/r = 0.37$: the system can be 
diagnosed as stable or unstable with reasonable confidence, 
but the precise value of $r$ is poorly resolved, with a 
relative uncertainty of $37\%$.  Furthermore, the inferred point lies \emph{below} the GOE 
boundary, signalling not a failure of inference but a breakdown 
of the GOE approximation itself: intracranial EEG dynamics are 
strongly non-normal, with directed, asymmetric connectivity 
between cortical regions, so that the appropriate ensemble is 
the Ginibre rather than the GOE. For the Ginibre ensemble, 
however, the OU process does not decouple into independent 
modes, since the eigenvectors of a non-symmetric matrix are 
not orthogonal; this prevents a mode-by-mode analysis of the 
effective sample size and precludes a closed-form analytical 
expression for the resolvability boundary. A full analytical treatment 
is left for future work.

\begin{figure}[t]
\centering
\includegraphics[width=0.65\textwidth]{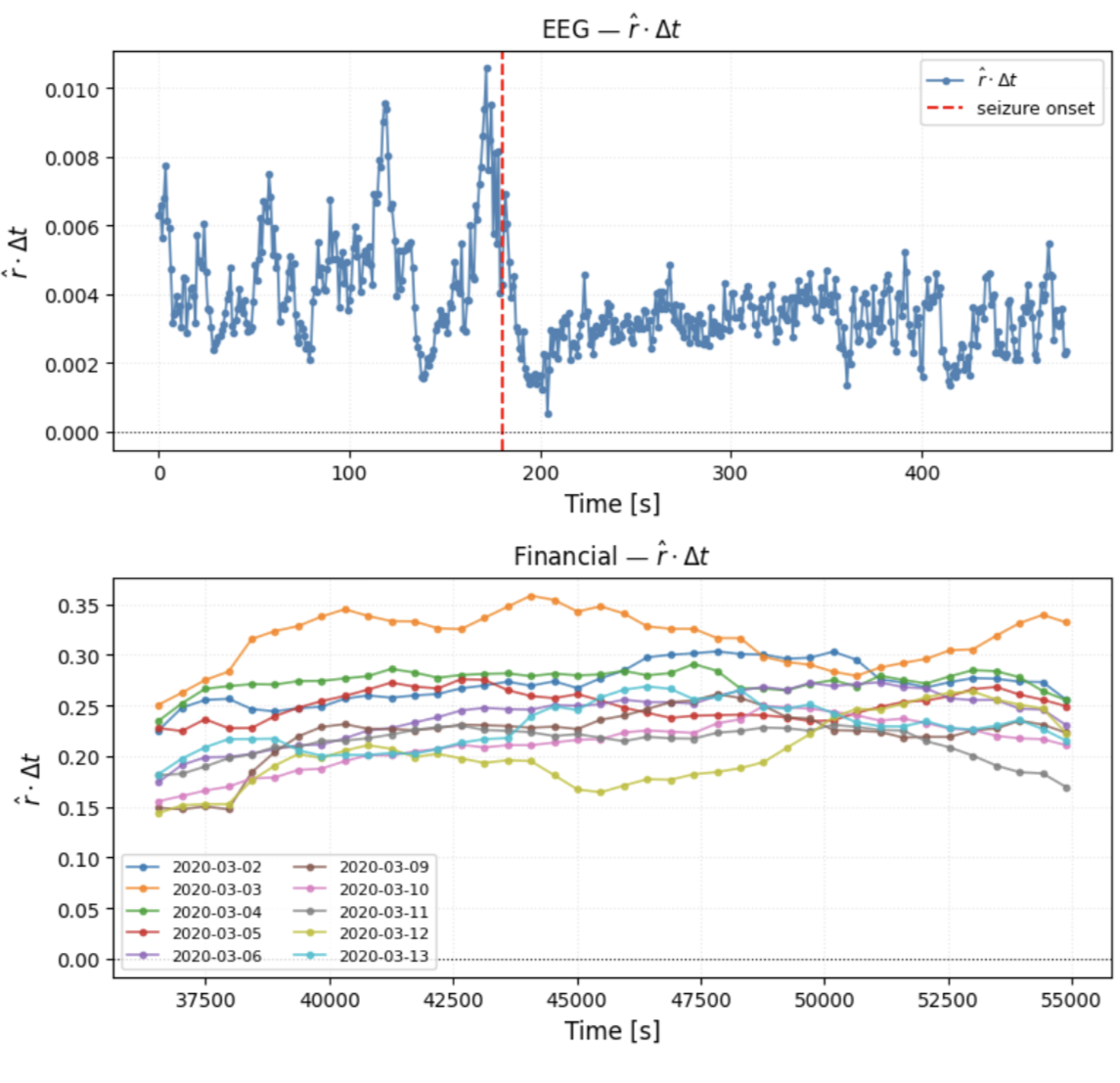}
\caption{
Estimated distance to instability $\hat r \cdot \Delta t$ across 
sliding windows for the EEG and financial datasets.
\emph{Top}: EEG recording from a representative patient. Each point 
corresponds to a sliding window estimate; the red dashed line marks 
seizure onset. In the pre-ictal phase $\hat r \cdot \Delta t$ 
exhibits large fluctuations, suggesting that the system repeatedly 
approaches and retreats from the stability boundary before 
eventually crossing it. At seizure onset the inferred distance 
drops sharply and remains suppressed throughout the post-ictal 
phase, consistent with a transient approach to instability during 
the ictal event.
\emph{Bottom}: Financial volatility panel ($N=47$ equities, 
$\Delta t = 1\,\text{s}$) across the ten trading days of 
2--13 March 2020. Each curve corresponds to one trading day; 
within-day non-overlapping windows are used to avoid overnight 
discontinuities. The inferred $\hat r \cdot \Delta t$ remains 
well above zero throughout, indicating robust distance from 
instability, with day-to-day variability reflecting changing 
market conditions during the COVID-19 onset period.
For the plankton dataset a single window spanning the full 
$T=264$ time points yields $\hat r \cdot \Delta t = 0.0196$.
}
\label{fig:rhat_timeseries}
\end{figure}

\newpage

\section{Discussion}\label{sec:discussion}

We have established a quantitative statement of a hypothesis common to
several fields: the precision with which the distance of a system from
instability can be estimated decreases as that distance decreases. As 
the slowest mode softens, its relaxation time diverges and consecutive 
observations become nearly indistinguishable, while the stationary 
fluctuations grow and obscure the deterministic signal. Together these 
two effects cause the curvature of the log-likelihood to vanish along 
the eigendirection that determines stability, so that the Fisher 
information loses its smallest eigenvalue at the threshold and 
inference becomes asymptotically uninformative.

A key insight is that the relevant sample size is not the nominal 
count $M$ but the effective count $M_{\rm eff}$, which accounts for 
temporal correlations and can fall far below $M$ near criticality. 
The compound-Wishart correction $\kappa_{\rm cW}$ captures this 
reduction and provides a good approximation of the true inflation 
factor, as confirmed by Monte Carlo simulations. When $M_{\rm eff}$ 
falls below the system dimension $N$, inference breaks down 
entirely, a threshold that is strictly more demanding than the 
naive requirement $M > N$, and that a near-critical system can 
violate even with an apparently adequate dataset.
Finite-size 
effects are accounted for through the Tracy--Widom correction 
to the expected distance to instability $\bar{r}(c,N)$, which 
shifts the resolvability boundary systematically toward higher 
stability for finite $N$.
The optimal sampling interval $\Delta t^* \approx 0.80/\bar{r}$ 
reflects a fundamental trade-off: consecutive samples separated 
by much less than the slowest relaxation time carry nearly 
identical information, whereas samples separated by much more 
discard dynamical content. The divergence of $\Delta t^*$ as 
$r \to 0$ implies that near-critical systems are hardest to 
sample efficiently, and that increasing raw data volume without 
adjusting the sampling rate yields diminishing inferential returns. 
In practice, the sampling interval should be matched to the 
system's correlation times, a constraint that becomes increasingly 
stringent as the system approaches the stability boundary.

It is also worth noting that, near criticality, relaxation 
times diverge and the linear approximation underlying the OU 
framework becomes progressively less accurate. In addition, the 
stationary assumption becomes increasingly difficult to justify 
over finite observation windows. The large uncertainty on $\hat{r}$ 
near the threshold is therefore not a limitation of the estimator 
but an intrinsic property of the problem.

The three empirical datasets span the full range of inferential 
regimes identified by the theory, and are characterised according 
to two criteria: whether the stability of the system can be 
reliably diagnosed, and whether the distance to instability can 
be quantitatively resolved. Financial volatility data meets both 
criteria, confirming that high-frequency equity data during 
market stress provides sufficient inferential power to locate 
the system relative to the stability boundary with high 
precision. Plankton abundance data meets neither: despite a nominally 
adequate sample size, temporal correlations render the observations 
effectively redundant and even the sign of $r$ cannot be 
reliably determined. Intracranial EEG data meets the diagnosis 
criterion but not the resolution criterion: the system is 
reliably diagnosed as stable, but the precise value of $r$ is 
not quantitatively resolved. Numerical results for the Ginibre ensemble 
are provided, and a closed-form analytical treatment is left 
for future work. More broadly, the framework opens several 
directions: extending the resolvability analysis to other 
matrix ensembles and spectral structures, accounting for the 
joint statistics of eigenvalues and eigenvectors to characterise 
mode mixing near criticality, and incorporating nonlinear 
dynamical models to understand inference in the regime where 
the linear stationary description breaks down.

\section{Methods}\label{sec:methods}

\textbf{Multivariate Ornstein--Uhlenbeck process.}
We model the system as a multivariate Ornstein--Uhlenbeck (OU) 
process,
\begin{equation}
d\mathbf{X}(t) = -\mathbf{A}\,\mathbf{X}(t)\,dt + \boldsymbol{\eta}(t),
\end{equation}
where $\mathbf{X}(t)\in\mathbb{R}^N$ is the state vector, 
$\mathbf{A}\in\mathbb{R}^{N\times N}$ is the drift matrix, and 
$\boldsymbol{\eta}(t)\in\mathbb{R}^N$ is Gaussian white noise with 
covariance $\langle\boldsymbol{\eta}(t)\boldsymbol{\eta}(s)^\top\rangle 
= 2\mathbf{B}\,\delta(t-s)$, where $\mathbf{B}$ is the diffusion 
matrix. The process is stable when all eigenvalues of $\mathbf{A}$ 
have positive real part; in this case it admits a unique stationary 
distribution $\mathbf{X}\sim\mathcal{N}(0,\boldsymbol{\Sigma}_\infty)$, 
where the stationary covariance $\boldsymbol{\Sigma}_\infty$ is well 
defined and satisfies the continuous Lyapunov equation
\begin{equation}
\mathbf{A}\boldsymbol{\Sigma}_\infty
+ \boldsymbol{\Sigma}_\infty\mathbf{A}^\top = 2\mathbf{B}.
\label{eq:lyapunov}
\end{equation}
The distance to instability is $r := \min_i\Re[\lambda_i(\mathbf{A})]$, 
which vanishes as the system approaches the critical point.\\

\textbf{Reversibility and Gaussian Orthogonal Ensemble.}
We focus on the reversible case, where the Onsager condition
\begin{equation}
    \mathbf{A}\mathbf{B} = \mathbf{B}\mathbf{A}^\top
\end{equation}
implies that
$\mathbf{A}$ and $\mathbf{B}$ are simultaneously diagonalisable.
Following~\cite{ferreira2025random}, a change of variables maps
the dynamics onto an equivalent system with $\mathbf{B} = \mathbf{I}$
and symmetric $\mathbf{A}$, for which the Lyapunov equation gives
$\boldsymbol{\Sigma}_\infty = \mathbf{A}^{-1}$. We work in this
basis throughout.\\

The natural ensemble for symmetric random matrices is the GOE,
and we model $\mathbf{A} = \mathbf{I} + \mathbf{W}$ with
$\mathbf{W} \sim \mathrm{GOE}(0, c^2/N)$. In the large-$N$ limit
the eigenvalue density converges to the Wigner semicircle shifted
to unit mean,
\begin{equation}
    \rho(\lambda) = \frac{1}{\pi c^2}
    \sqrt{2c^2 - (\lambda-1)^2}\;
    \mathbf{1}_{\{|\lambda - 1| \leq \sqrt{2}\,c\}},
\end{equation}
with stability requiring $c < c_{\rm crit} = 1/\sqrt{2}$ and
distance to instability $r = 1 - \sqrt{2}\,c$.At finite $N$, the smallest eigenvalue fluctuates according to the
Tracy--Widom$_1$ law~\cite{tw2}, yielding the finite-size
corrected distance
\begin{equation}
    \bar{r}(c,N)
    =
    (1-\sqrt{2}\,c)
    -
    (\sqrt{2}\,c)^{1/3}\,\mu_{\mathrm{TW}}\,N^{-2/3},
    \qquad
    \mu_{\mathrm{TW}}\approx -1.21,
    \label{eq:rbar}
\end{equation}
so that finite systems are on average slightly more stable than
predicted by the large-$N$ limit, although a finite fraction of
realizations may still cross the stability threshold. The restriction
to symmetric $\mathbf{A}$ is not essential: qualitatively identical
results hold for asymmetric matrices with the GOE replaced by the
Ginibre ensemble.\\

\textbf{Discrete-time representation and Bayesian inference.}
We observe a discrete trajectory 
$\mathbf{X} = (\mathbf{X}_0, \mathbf{X}_1, \dots, \mathbf{X}_{M-1})$ 
at equally spaced times with sampling interval $\Delta t$. 
Integrating the OU process over one time step, the continuous-time 
dynamics reduce exactly to a VAR(1) process,
\begin{equation}
\mathbf{X}_{k+1} = \mathbf{Q}\mathbf{X}_k + \boldsymbol{\varepsilon}_k,
\qquad
\boldsymbol{\varepsilon}_k \sim \mathcal{N}(0, \boldsymbol{\Sigma}_{\Delta t}),
\end{equation}
where $\mathbf{Q} = e^{-\mathbf{A}\Delta t}$ is the discrete 
propagator and $\boldsymbol{\Sigma}_{\Delta t} = 
\boldsymbol{\Sigma}_\infty - \mathbf{Q}\boldsymbol{\Sigma}_\infty
\mathbf{Q}^\top$ is the innovation covariance. The discretely 
sampled process is therefore fully characterised by the parameter 
set $\boldsymbol{\theta} = (\mathbf{Q}, \boldsymbol{\Sigma}_{\Delta t})$, 
which is directly accessible from the observed data.

We adopt a Bayesian framework with uniform prior 
$P(\boldsymbol{\theta}) \propto \mathrm{const}$, so that the 
posterior
\begin{equation}
P(\boldsymbol{\theta} \mid \mathbf{X})
\propto P(\mathbf{X} \mid \boldsymbol{\theta})
\end{equation}
is entirely determined by the likelihood. Conditioning on the 
initial state $\mathbf{X}_0$ and exploiting the Markov property, 
the log-likelihood reads
\begin{equation}
\begin{aligned}
\log P(\mathbf{X} \mid \boldsymbol{\theta})
&= -\frac{1}{2}\log\det(2\pi\boldsymbol{\Sigma}_\infty)
   -\frac{1}{2}\mathbf{X}_0^\top \boldsymbol{\Sigma}_\infty^{-1} 
   \mathbf{X}_0 \\
&\quad -\frac{M-1}{2}\log\det(2\pi\boldsymbol{\Sigma}_{\Delta t})
   - \frac{1}{2}\sum_{k=0}^{M-2}
(\mathbf{X}_{k+1} - \mathbf{Q}\mathbf{X}_k)^\top
\boldsymbol{\Sigma}_{\Delta t}^{-1}
(\mathbf{X}_{k+1} - \mathbf{Q}\mathbf{X}_k).
\end{aligned}
\end{equation}

Via the trace identity $\mathbf{x}^\top \mathbf{A} \mathbf{y} = 
\mathrm{Tr}(\mathbf{A}\,\mathbf{y}\mathbf{x}^\top)$, the quadratic 
form reduces to a function of three sufficient statistics \cite{singh2017fast},
\begin{equation}
T_1 = \frac{1}{M-1}\sum_{k=1}^{M-1}\mathbf{X}_k\mathbf{X}_k^\top,
\qquad
T_2 = \frac{1}{M-1}\sum_{k=1}^{M-1}\mathbf{X}_k\mathbf{X}_{k-1}^\top,
\qquad
T_3 = \frac{1}{M-1}\sum_{k=0}^{M-2}\mathbf{X}_k\mathbf{X}_k^\top.
\end{equation}

Completing the square in $\mathbf{Q}$ and optimising over 
$\boldsymbol{\Sigma}_{\Delta t}$ yields the maximum a posteriori 
(MAP) estimators, which under the uniform prior coincide with 
the maximum-likelihood estimators,
\begin{equation}
\hat{\mathbf{Q}} = T_2 T_3^{-1},
\qquad
\hat{\boldsymbol{\Sigma}}_{\Delta t} = T_1 - T_2 T_3^{-1} T_2^\top.
\end{equation}

The continuous-time parameters are then recovered as
\begin{equation}
\hat{\mathbf{A}} = -\frac{1}{\Delta t}\log\hat{\mathbf{Q}},
\qquad
\hat{\mathbf{B}} = \frac{1}{2}\!\left(
\hat{\mathbf{A}}\,\hat{\boldsymbol{\Sigma}}_\infty
+ \hat{\boldsymbol{\Sigma}}_\infty\,\hat{\mathbf{A}}^\top
\right),
\end{equation}
where $\hat{\boldsymbol{\Sigma}}_\infty$ solves the discrete 
Lyapunov equation
\begin{equation}
\hat{\boldsymbol{\Sigma}}_\infty =
\hat{\mathbf{Q}}\,\hat{\boldsymbol{\Sigma}}_\infty\hat{\mathbf{Q}}^\top
+ \hat{\boldsymbol{\Sigma}}_{\Delta t}.
\end{equation}
The estimators are consistent: $\hat{\mathbf{Q}} \to \mathbf{Q}$ 
and $\hat{\mathbf{A}} \to \mathbf{A}$ as $M \to \infty$. The 
full derivation is given in Supplementary Section~1.\\

\textbf{Fisher information matrix.}
The precision of parameter estimation is governed by the Fisher
information matrix, defined as the expected negative Hessian of
the log-likelihood,
\begin{equation}
    \mathcal{I}(\boldsymbol{\theta}_\star)
    = -\mathbb{E}_{\mathbf{X}\mid\boldsymbol{\theta}_\star}\!\left[
    \nabla^2_{\boldsymbol{\theta}}
    \log P(\mathbf{X}\mid\boldsymbol{\theta})
    \Big|_{\boldsymbol{\theta}=\boldsymbol{\theta}_\star}\right],
\end{equation}
which sets the Cram\'er--Rao lower bound $\mathrm{Cov}(\hat{\boldsymbol{\theta}})
\geq \mathcal{I}^{-1}$ on the variance of any unbiased estimator.
The Fisher information matrix is block-diagonal with blocks
\begin{equation}
\mathcal{I}^{(\Sigma\Sigma)}
= \boldsymbol{\Sigma}_{\Delta t} \otimes \boldsymbol{\Sigma}_{\Delta t},
\qquad
\mathcal{I}^{(QQ)}
= 2\,\mathbf{J} \otimes \boldsymbol{\Sigma}_\infty.
\end{equation}
The block $\mathcal{I}^{(QQ)}$ governs the uncertainty on the
distance to instability;
Since the true parameters are unknown in practice, we use the
plug-in estimator $\hat{\mathcal{I}} = \mathcal{I}(\hat{\boldsymbol{\Sigma}}_{\Delta t},
\hat{\boldsymbol{\Sigma}}_\infty)$. In our simulations we perform
a double Monte Carlo: an outer loop over $n_A$ realisations of
$\mathbf{A}$ from the GOE, and an inner loop over $n_{\rm sim}$
trajectories per realisation. For each realisation we compute
the theoretical $\mathcal{I}(\boldsymbol{\theta}_\star)$ and the
plug-in estimator $\hat{\mathcal{I}}$, obtained by averaging the
empirical Hessian over trajectories consistently with
$\mathcal{I} = \mathbb{E}_{\mathbf{X}\mid\mathbf{Q}_\star}[\mathbf{H}]$,
retaining only realisations with $\lambda_{\min}(\hat{\mathbf{A}})
\geq 0$.

The unconditional covariance of $\hat{\mathbf{Q}}$ acquires an
inflation factor $\kappa_W \geq 1$ from the convexity of matrix
inversion (Jensen's inequality, $\mathbb{E}[T_3^{-1}] \geq
\boldsymbol{\Sigma}_\infty^{-1}$), so that
\begin{equation}
\mathrm{Cov}(\mathrm{vec}\,\hat{\mathbf{Q}})
= \frac{\kappa_W}{M-1}
\bigl(\boldsymbol{\Sigma}_\infty^{-1} \otimes
\boldsymbol{\Sigma}_{\Delta t}\bigr).
\label{eq:cov_kW}
\end{equation}
The derivation of $\kappa_W$ is given in Supplementary Section~2.\\

\textbf{Wishart inflation factor.}
When the regressors $\mathbf{X}_k$ were i.i.d.\ $\mathcal{N}(0,
\boldsymbol{\Sigma}_\infty)$, the sample covariance $(M-1)T_3$
follows a Wishart distribution $\mathcal{W}_N(M-1,
\boldsymbol{\Sigma}_\infty)$, and the inverse moment is exactly
\begin{equation}
\mathbb{E}\!\left[(M-1)^{-1}T_3^{-1}\right]
= \frac{\boldsymbol{\Sigma}_\infty^{-1}}{M-1 - N - 1}
= \frac{\boldsymbol{\Sigma}_\infty^{-1}}{M - N - 2},
\end{equation}
giving the inflation factor
\begin{equation}
    \kappa_W = \frac{M-1}{M-N-2}.
\end{equation}

\textbf{Effective sample size.}
When $\mathbf{A}$ is symmetric, the VAR(1) system decouples into
$N$ independent scalar AR(1) processes, one per eigenvector of
$\mathbf{A}$. For a single AR(1) process with autoregressive
coefficient $e^{-\lambda_\alpha \Delta t}$, the integrated
autocorrelation time of the sample variance is given by
Isserlis' theorem as
\begin{equation}
\tau_\alpha
= \frac{1 + e^{-2\lambda_\alpha \Delta t}}
       {1 - e^{-2\lambda_\alpha \Delta t}},
\end{equation}
which diverges as $\tau_\alpha \approx (\lambda_\alpha \Delta
t)^{-1}$ in the slow-mode limit $\lambda_\alpha \Delta t \ll 1$.
The $M$ observations along mode $\alpha$ are therefore equivalent
to
\begin{equation}
M_{\rm eff}^\alpha = \frac{M}{\tau_\alpha},
\end{equation}
independent samples, which vanishes as $M_{\rm eff}^\alpha
\approx M\lambda_\alpha\Delta t$ in the slow-mode limit.

Since $\lambda_{\min}(\mathbf{A})$ is extracted from the full
estimated matrix $\hat{\mathbf{A}}$, the relevant effective
sample size is a single global quantity. Slow modes contribute
disproportionately to the reduction of the inferential budget,
so we weight the autocorrelation times by their own magnitude,
obtaining the variance-to-mean ratio
\begin{equation}
M_{\rm eff}
\approx \frac{M}{\langle\tau\rangle_{\rm w}},
\qquad
\langle\tau\rangle_{\rm w}(c, \Delta t)
= \frac{\sum_{\alpha=1}^N \tau(\lambda_\alpha)^2}
       {\sum_{\alpha=1}^N \tau(\lambda_\alpha)},
\end{equation}
where $\tau(\lambda_\alpha)$ is defined above, and
$\langle\tau\rangle_{\rm w}$ depends on $c$ through the
eigenvalues $\{\lambda_\alpha\}$ of $\mathbf{A}$, which is
sampled from an ensemble with coupling strength $c$, and on
$\Delta t$ through $\tau_\alpha$.

A lower bound is obtained by retaining only the contribution
of the soft mode,
\begin{equation}
M_{\rm eff}^{\rm min} = \frac{M}{\tau_r} \approx Mr\Delta t,
\qquad \tau_r = \tau_{\max} \geq \langle\tau\rangle_{\rm w},
\end{equation}
where $r = \lambda_{\min}(\mathbf{A})$. The ratio $q_{\rm eff}
= M_{\rm eff}/N$ replaces the nominal $q = M/N$ as the relevant
control parameter: inference is feasible only when $q_{\rm eff}
> 1$, and a near-critical system can violate this condition even
when $q \gg 1$, because temporal correlations render the
observations effectively redundant.\\
\textbf{Compound-Wishart inflation factor.}
In the OU process, consecutive observations are temporally
correlated and $T_3$ is no longer a standard Wishart matrix.
A naive correction is obtained by replacing $M$ with
$M_{\rm eff}$, giving
\begin{equation}
    \kappa_{\rm cW}
    \approx \frac{M_{\rm eff} - 1}
           {M_{\rm eff} - N - 2},
    \label{eq:kcW}
\end{equation}
defined for $M_{\rm eff} > N+2$, and an upper bound using
$M_{\rm eff}^{\rm min}$,
\begin{equation}
    \kappa_{\rm cW}^{\rm max}
    = \frac{M/\tau_r - 1}
           {M/\tau_r - N - 2}.
\end{equation}
Both diverge at the breakdown threshold $M_{\rm eff} \to (N+2)^+$
and recover $\kappa_W$ as $\Delta t \to \infty$. Numerical
simulations confirm
\begin{equation}
    \kappa_W \leq \kappa_{\rm cW}
    \leq \kappa_{\rm cW}^{\rm max},
\end{equation}
with MC estimates lying within the error bars of $\kappa_{\rm cW}$
(figure~\ref{fig:projected_soft_mode_uncertainty}).

\paragraph{Variance of the distance to instability}
\label{sec:vardist2}
Projecting the covariance \eqref{eq:cov_kW} onto the soft 
eigenvector $\mathbf{u}_a$, the variance of the projected 
propagator estimate $\hat{q}_a = \mathbf{u}_a^\top 
\hat{\mathbf{Q}} \mathbf{u}_a$ is
\begin{equation}
    \mathrm{Var}(\hat{q}_a)
    = \frac{\kappa_{\rm cW}}{M}\,(1 - e^{-2\lambda_a\Delta t}).
\end{equation}
 Propagating through
$\hat{\lambda}_a = -\Delta t^{-1}\log\hat{q}_a$ via the delta
method,
\begin{equation}
    \mathrm{Var}(\hat{\lambda}_a)
    \approx \left(\frac{d\lambda_a}{dq_a}\right)^2
    \mathrm{Var}(\hat{q}_a)
    = \frac{\kappa_{\rm cW}}{M}\cdot
    \frac{e^{2\lambda_a\Delta t}-1}{\Delta t^2}.
\end{equation}

Specialising to the soft mode $\lambda_a = r$ and dividing by $r^2$,
the relative variance of the estimated distance to instability is
\begin{equation}
    \frac{\mathrm{Var}(\hat{r})}{r^2}
    = \frac{\kappa_{\rm cW}}{M}\cdot
    \frac{e^{2r\Delta t}-1}{r^2\Delta t^2}.
    \label{eq:rel_var_r}
\end{equation}
In the slow-mode limit $r\Delta t \ll 1$, which is precisely the
regime of interest near criticality, $e^{2r\Delta t}-1 \approx
2r\Delta t$ and the expression simplifies to
\begin{equation}
    \frac{\sigma_r}{r}
    \approx \sqrt{\frac{2\kappa_{\rm cW}}{M r\Delta t}},
    \label{eq:rel_var_approx}
\end{equation}
where $Mr\Delta t$ counts the total number of relaxation times
contained in the trajectory. The divergence of $\sigma_r/r$ as $r \to 0$ has two contributing 
sources. In the resolvable regime $M > (N+2)\langle\tau\rangle_{\rm w}$, 
$\kappa_{\rm cW}$ is finite and \eqref{eq:rel_var_approx} diverges 
as $1/\sqrt{r}$. In the breakdown regime $M \to 
(N+2)\langle\tau\rangle_{\rm w}$, $\kappa_{\rm cW} \to \infty$ 
independently of $r$, contributing an additional divergence.

To attain relative precision $\sigma_r/r \leq \varepsilon$ 
at fixed $\Delta t$, the number of observations must satisfy
\begin{equation}
    M \geq \frac{2\kappa_{\rm cW}}{\varepsilon^2\, r\, \Delta t},
    \label{eq:uncertainty}
\end{equation}
which diverges as $r \to 0$ at any fixed target precision 
$\varepsilon$: the closer the system is to instability, the 
more observations are required to achieve the same relative 
precision.\\

\textbf{Resolvability boundaries and optimal sampling.}
A direct consequence of \eqref{eq:rel_var_approx} is the 
probability of a false instability diagnosis. Under the Gaussian 
approximation $\hat{r}\approx\mathcal{N}(r,\sigma_r^2)$, the 
probability that the estimated distance to instability is 
negative, i.e.\ that a stable system is falsely declared 
unstable, is
\begin{equation}
    P_{\mathrm{false}}
    = \Pr(\hat{r} < 0)
    = \Phi\!\left(-\frac{r}{\sigma_r}\right)
    = \Phi\!\left(-\frac{1}{\sigma_r/r}\right),
    \label{eq:Pfalse}
\end{equation}
where $\Phi$ is the standard normal CDF. Two thresholds are 
considered. The stability diagnosis threshold $\sigma_r/r = 1$ 
corresponds to $P_{\mathrm{false}} = \Phi(-1)\approx 16\%$: 
more than one in six estimates incorrectly diagnoses instability. 
The resolution threshold $\sigma_r/r = 0.1$ corresponds to 
$P_{\mathrm{false}} < 0.01\%$, providing a stringent criterion 
for quantitative precision.

Combining \eqref{eq:rel_var_approx} with a prescribed threshold 
$\sigma_r/r = \varepsilon$, the resolvability boundary in the 
$(c,q,\Delta t)$ parameter space is defined by
\begin{equation}
    \frac{\sigma_r}{r}(c,q,\Delta t,N)
    = \sqrt{\frac{2\kappa_{\rm cW}}{\bar{r}\,q\,N\,\Delta t}} 
    = \varepsilon,
    \label{eq:boundary}
\end{equation}
with $\bar{r} = \bar{r}(c,N)$ from \eqref{eq:rbar_} and the 
slow-mode approximation $\bar{r}\Delta t \ll 1$. Setting 
$\varepsilon = 1$ gives the diagnosis boundary; setting 
$\varepsilon = 0.1$ gives the resolution boundary.

At fixed number of samples $M$, the relative variance
\eqref{eq:rel_var_r} depends on $\Delta t$ through the factor
$(e^{2r\Delta t}-1)/\Delta t^2$. Setting $u = 2r\Delta t$ and
minimising $(e^u-1)/u^2$ over $u > 0$,
\begin{equation}
    \frac{d}{du}\frac{e^u-1}{u^2}
    = \frac{ue^u - 2(e^u-1)}{u^3} = 0
    \quad\Longrightarrow\quad
    e^u(2-u) = 2,
    \label{eq:opt_dt_eq}
\end{equation}
which is solved numerically by $u^*\approx 1.59$, giving
\begin{equation}
    \Delta t^* = \frac{u^*}{2\bar{r}} \approx \frac{0.80}{\bar{r}}.
    \label{eq:opt_dt}
\end{equation}
Sampling at $\Delta t < \Delta t^*$ reduces the total observation 
time $M\Delta t$ without a compensating gain in information per 
sample; sampling at $\Delta t > \Delta t^*$ increases the variance 
penalty $(e^{2r\Delta t}-1)/\Delta t^2$. The optimum diverges as 
$\bar{r}\to 0$, confirming that near-critical systems require 
increasingly slow sampling to extract independent information.

\textbf{Frobenius error and criticality.}
It is tempting to measure estimation quality by the matrix
reconstruction error $\|\hat{\bm{A}} - \bm{A}\|_F^2$. This
quantity is \emph{blind} to criticality in the resolvable regime,
which sharpens what the uncertainty on $r$ is about: resolving
the distance to instability, not reconstructing the full matrix
$\bm{A}$.

In the reversible case, $\bm{A}$ and $\hat{\bm{A}}$ are both
symmetric. Assuming eigenvector alignment to leading order, the
squared Frobenius error reduces to
\begin{equation}
    \|\hat{\bm{A}} - \bm{A}\|_F^2
    = \sum_a (\hat{\lambda}_a - \lambda_a)^2,
    \label{eq:frob_decomp}
\end{equation}
i.e.\ the sum of squared eigenvalue errors (see Supplemental section 7). Taking the expectation
over the observed trajectory $\bm{X}$,
\begin{equation}
    \mathbb{E}_{\bm{X}|\mathbf{Q}_*}\!\left[\sum_a(\hat{\lambda}_a -
    \lambda_a)^2\right]
    = \sum_a \mathrm{Var}_{\bm{X}|\mathbf{Q}_*}(\hat{\lambda}_a)
    = \frac{\kappa_{\rm cW}}{M\Delta t^2}
    \sum_a (e^{2\lambda_a\Delta t} - 1).
    \label{eq:frob_diag}
\end{equation}
Since $(e^{2\lambda_a\Delta t}-1)$ grows with $\lambda_a$, the sum
is dominated by the stiff modes. In the resolvable regime
$M_{\rm eff} > N+2$, the soft mode contributes
\begin{equation}
    \frac{\kappa_{\rm cW}}{M\Delta t^2}(e^{2r\Delta t} - 1)
    \xrightarrow{\;r\to 0\;} 0,
    \label{eq:frob_soft}
\end{equation}
since $(e^{2r\Delta t}-1) \approx 2r\Delta t \to 0$ linearly in
$r$ while $\kappa_{\rm cW} \to \kappa_W$ remains finite. The
total Frobenius error therefore stays $O(1)$ and is blind to the
approach to criticality: the stiff modes dominate and mask the
divergence of $\sigma_r/r$.

This blindness breaks down at the boundary of the resolvable
regime. When $M_{\rm eff} \to (N+2)^+$, $\kappa_{\rm cW}$
diverges and so does the Frobenius error. The two failure modes
are therefore distinct: $\sigma_r/r$ grows continuously as
$r \to 0$ and can become large well before the breakdown
threshold is reached, while the Frobenius error remains $O(1)$
throughout the resolvable regime and diverges only at the
breakdown threshold $M_{\rm eff} = N+2$, where both quantities
fail simultaneously.

\begin{figure}[t]
\centering
\includegraphics[width=0.95\textwidth]{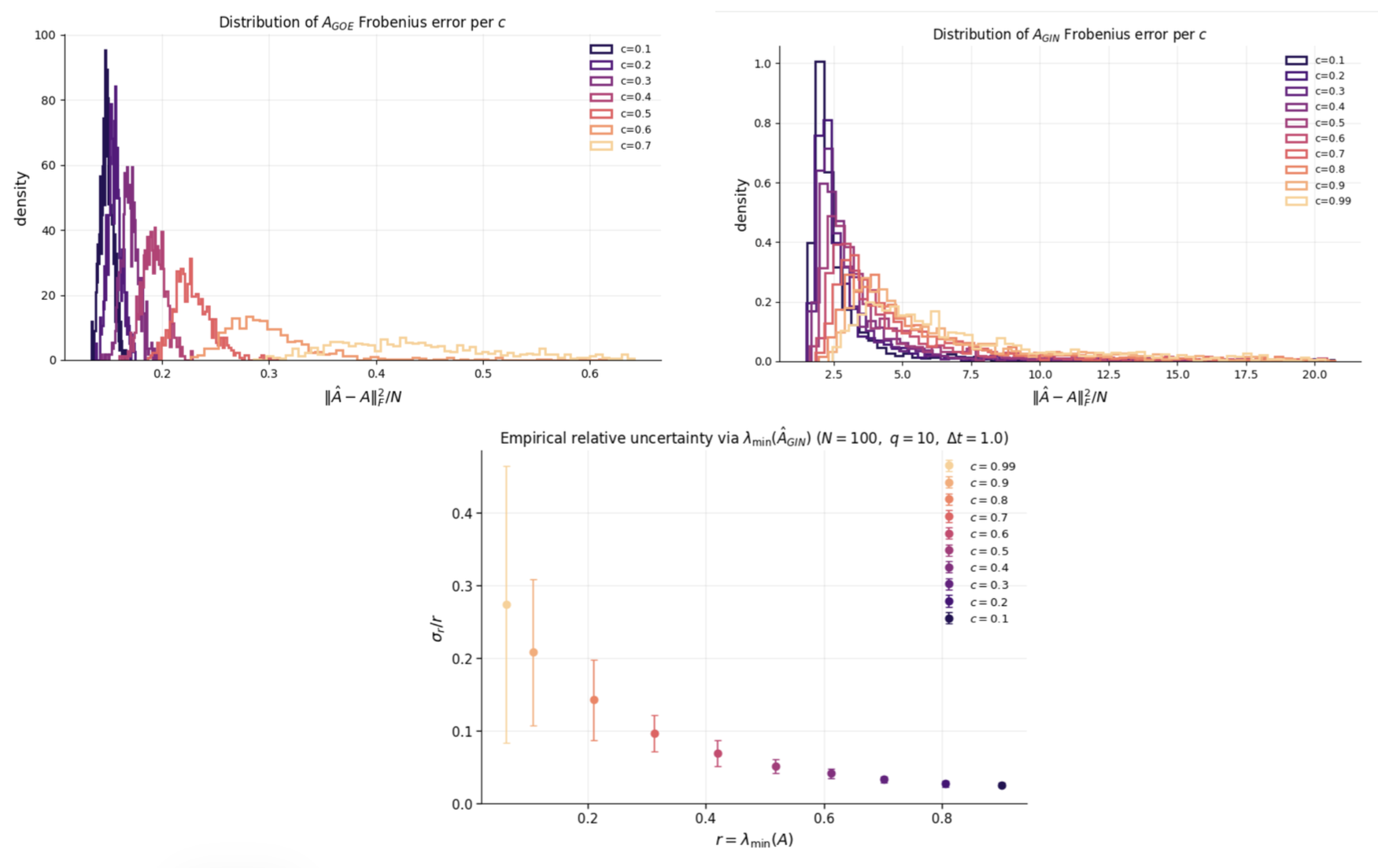}
\caption{
\emph{Top left}: Distribution of the Frobenius reconstruction
error $\|\hat{\mathbf{A}}-\mathbf{A}\|_F^2/N$ for the GOE
ensemble ($N=100$, $q=40$, $\Delta t=1$), for coupling strengths
$c\in\{0.1,\ldots,0.7\}$. As $c$ increases toward
$c_{\rm crit}=1/\sqrt{2}$, the distribution broadens but the
mean remains finite, confirming that the Frobenius error is
blind to criticality: the stiff modes dominate the reconstruction
error and mask the divergence of the soft mode.
\emph{Top right}: Same quantity for the Ginibre ensemble
($N=100$, $q=10$, $\Delta t=1$), for $c\in\{0.1,\ldots,0.99\}$.
The errors are larger than in the GOE case due to the smaller
observation ratio $q$ and the asymmetric structure of the
Ginibre drift matrix, but again remain finite as
$c \to c_{\rm crit} = 1$.
\emph{Bottom}: Empirical relative uncertainty $\sigma_r/r$
for the Ginibre ensemble, estimated from
$\lambda_{\min}(\hat{\mathbf{A}})$ across Monte Carlo
realisations. Each point corresponds to one value of $c$,
colored by coupling strength. No theoretical prediction is
available for the Ginibre ensemble. As $c$ approaches
$c_{\rm crit} = 1$ from below, $\sigma_r/r$ grows by more
than an order of magnitude, signalling that the estimated
distance carries a large relative uncertainty and the precise
value of $r$ can no longer be reliably resolved. This is
consistent with the fundamental limit established analytically
for the GOE: the signal vanishes faster than the noise as the
system approaches the stability threshold, regardless of the
symmetry class of the coupling matrix.}
\label{fig:frob_err}
\end{figure}

\textbf{Empirical datasets.}
We apply our inference procedure to three multivariate time
series from qualitatively distinct domains, all modelled as
OU processes and analysed with the same estimation pipeline.
The first dataset consists of intraday log-returns of $N=47$
large-cap US equities during early March 2020, coinciding with
the onset of the COVID-19 pandemic ($\Delta t = 1\,$s);
absolute log-returns serve as a proxy for volatility and the
analysis is performed on non-overlapping within-day windows
to avoid overnight gaps. The second is the coastal bacterial
abundance time series of Martin-Platero et al.\
\cite{martinplatero2018plankton}, retaining the $N=47$ most abundant
OTUs across $T=264$ time points sampled three times daily over
88 consecutive days ($\Delta t = 1/3\,$d); abundances are
log-transformed, centred per species, and rescaled by a global
standard deviation. The third comprises intracranial EEG
recordings from 16 epileptic patients; here we focus on a
representative single-patient recording structured as a preictal
segment (3 min), an ictal segment of variable duration, and a
postictal segment (3 min), sampled at $f_s = 512\,$Hz across
$N=47$ electrodes ($\Delta t = 1/512 \approx 0.002\,$s), with
each channel centred and rescaled by a global standard deviation.

\newpage

\section*{Declarations}
\begin{itemize}
\item The authors declare no competing interests.
\end{itemize}

\bibliography{RMT.bib}

\newpage

\appendix

\begin{center}
    {\bf \Large Supplementary Information: \\ Fundamental limits of stability inference \\ in high-dimensional complex systems}
\end{center}
\begin{abstract}
\noindent This Supplementary Information collects the full derivations
summarised in the main text and extends them in several directions. We
derive the complete log-likelihood for the discretely sampled multivariate
Ornstein--Uhlenbeck process, the maximum-likelihood estimators, and their
sufficient statistics. We show that the estimator's sampling covariance is
exactly the inverse Hessian of the log-likelihood, and identify two
multiplicative inflation factors above the information-theoretic floor:
the inverse-Wishart factor arising from substituting the noisy sample
covariance for the true one, and a compound-Wishart correction that
accounts for the temporal autocorrelation of the data. Propagating this
variance to the eigenvalues of the estimated drift matrix yields the
soft-mode law, whose divergence as the distance to instability vanishes
is the quantitative expression of critical slowing down of inference and
implies a hard uncertainty bound that no estimator can beat. We show that
the Frobenius reconstruction error is blind to this divergence: it is
dominated by the stiff modes and remains bounded at criticality. The
finite-size distribution of the distance to instability is governed by
Tracy-Widom edge statistics, which provide the floor used to construct
the resolvability phase diagram of the main text. Finally, we report
numerical experiments on three empirical time series: a representative
single-patient intracranial EEG recording across seizure onset, intraday
financial volatility during the COVID-19 market dislocation of March 2020,
and a coastal bacterial abundance series. All three are modelled as
multivariate Ornstein-Uhlenbeck processes and analysed with the same
estimation pipeline.
\end{abstract}

\newpage
\tableofcontents
\newpage


\section{Model and assumptions}\label{sec:model}
\subsection{Multivariate Ornstein--Uhlenbeck}
We model the system as a multivariate Ornstein--Uhlenbeck (MVOU) process
$\mathbf{X}(t)\in\mathbb{R}^N$ solving the It\^o stochastic differential
equation
\begin{equation}
    d\mathbf{X}(t) = -\mathbf{A}\,\mathbf{X}(t)\,dt + \boldsymbol{\eta}(t),
    \label{eq:sde}
\end{equation}
where $\mathbf{A}\in\mathbb{R}^{N\times N}$ is the drift matrix, $\boldsymbol{\eta}(t)\in\mathbb{R}^N$ is Gaussian
white noise with covariance
\begin{equation}
    \mathbb{E}\bigl[\boldsymbol{\eta}(t)\,\boldsymbol{\eta}(t')^\top\bigr]
    = 2\mathbf{B}\,\delta(t-t').
    \label{eq:noise_cov}
\end{equation}
and $\mathbf{B}\in\mathbb{R}^{N\times N}$ is the symmetric positive-definite
diffusion matrix.\\
The process is stable when all eigenvalues of $\mathbf{A}$ have strictly
positive real part; the distance to instability is
\begin{equation}
    r := \min_i\operatorname{Re}[\lambda_i(\mathbf{A})] > 0.
\end{equation}
When stable, the process admits a unique stationary Gaussian distribution
$\mathcal{N}(0,\boldsymbol{\Sigma}_\infty)$, where the stationary
covariance $\boldsymbol{\Sigma}_\infty\in\mathbb{R}^{N\times N}$
satisfies the Lyapunov equation
\begin{equation}
    \mathbf{A}\boldsymbol{\Sigma}_\infty
    + \boldsymbol{\Sigma}_\infty\mathbf{A}^\top = 2\mathbf{B}.
    \label{eq:lyapunov_}
\end{equation}
As $r\to 0^+$ the system approaches instability and the stationary approximation
breaks down.
The following assumptions are maintained throughout. The data are
generated exactly by \eqref{eq:sde} with $\mathbf{A}$ and $\mathbf{B}$
constant over the observation window. The observed trajectory
$\{\mathbf{X}_k\}_{k=0}^{M-1}$ is a single realisation of the
stationary process, so that $\mathbf{X}_0 \sim
\mathcal{N}(0,\boldsymbol{\Sigma}_\infty)$ and the process is assumed to
have reached stationarity prior to $k=0$. Observations are equally
spaced with interval $\Delta t > 0$, and the number of observed
transitions $M-1$ satisfies $M - 1 > N$, i.e.\ the observation ratio
$q := (M-1)/N > 1$, which is the necessary and sufficient condition for
$T_3$ to be invertible almost surely and for the maximum-likelihood estimators to be well defined.

\paragraph{Reversibility.}
We focus primarily on the reversible case, defined by the
Onsager reversibility condition
\begin{equation}
    \mathbf{A}\mathbf{B} = \mathbf{B}\mathbf{A}^\top,
    \label{eq:reversibility}
\end{equation}
which implies that $\mathbf{A}$ and $\mathbf{B}$ are simultaneously
diagonalisable and that the stationary process satisfies detailed
balance. Following~\cite{ferreira2025random}, we perform the change of
variables $\mathbf{X}' = \mathbf{B}^{-1}\mathbf{X}$,
which maps the dynamics onto an equivalent system with diffusion matrix
$\mathbf{B}' = \mathbf{I}_N$ and symmetric drift matrix
$\mathbf{A}' = \mathbf{B}^{-1}\mathbf{A}\mathbf{B}$,
for which the Lyapunov equation~\eqref{eq:lyapunov_} gives immediately
\begin{equation}
    \boldsymbol{\Sigma}'_\infty = (\mathbf{A}')^{-1}.
    \label{eq:sigma_transformed}
\end{equation}
Writing the spectral decomposition $\mathbf{A}' =
\mathbf{U}\,\mathrm{diag}(\lambda_1,\dots,\lambda_N)\,\mathbf{U}^\top$
with $\mathbf{U}\in\mathbb{R}^{N\times N}$ orthogonal, the rotated
coordinates $x_a := (\mathbf{U}^\top\mathbf{X}')_a$ decouple into $N$
independent scalar AR(1) processes,
\begin{equation}
    x_{a,k+1} = q_a\, x_{a,k} + \eta_{a,k},
    \qquad
    q_a = e^{-\lambda_a \Delta t},
    \qquad
    \eta_{a,k} \sim \mathcal{N}\!\left(0,\,\frac{1-q_a^2}{\lambda_a}\right),
    \label{eq:scalar_modes}
\end{equation}
where $s_a := 1/\lambda_a$ is the stationary variance of mode $a$ in
the transformed system.

\subsection{Exact VAR(1) discretisation}
We observe $\mathbf{X}_k := \mathbf{X}(k\Delta t)$ for $k=0,\dots,M-1$.
Integrating \eqref{eq:sde} over one step by the variation-of-constants
formula gives
\begin{equation}
    \mathbf{X}(t+\Delta t)
    = e^{-\mathbf{A}\Delta t}\mathbf{X}(t)
    + \int_0^{\Delta t} e^{-\mathbf{A}(\Delta t-s)}\,\boldsymbol{\eta}(t+s)\,ds,
\end{equation}
which at discrete times $t = k\Delta t$ yields the VAR(1)
representation 
\begin{equation}
    \mathbf{X}_{k+1} = \mathbf{Q}\,\mathbf{X}_k + \boldsymbol{\varepsilon}_k,
    \label{eq:var1}
\end{equation}
with $\mathbf{Q} = e^{-\mathbf{A}\Delta t}$
and innovation
\begin{equation}
    \boldsymbol{\varepsilon}_k
    = \int_0^{\Delta t} e^{-\mathbf{A}(\Delta t-s)}\,\boldsymbol{\eta}(k\Delta t+s)\,ds.
    \label{eq:eps_def}
\end{equation}
Since $\boldsymbol{\eta}$ is white noise independent of
$\mathbf{X}(k\Delta t)$, the innovation $\boldsymbol{\varepsilon}_k$
is Gaussian, independent of $\mathbf{X}_k$, and mutually independent
across $k$. Its covariance is
\begin{align}
    \boldsymbol{\Sigma}_{\Delta t}
    &= \mathbb{E}\!\left[\boldsymbol{\varepsilon}_k\boldsymbol{\varepsilon}_k^\top\right]
    \nonumber\\
    &= \int_0^{\Delta t}\int_0^{\Delta t}
    e^{-\mathbf{A}(\Delta t-s)}\,
    \mathbb{E}\!\left[\boldsymbol{\eta}(k\Delta t+s)\,
    \boldsymbol{\eta}(k\Delta t+s')^\top\right]
    e^{-\mathbf{A}^\top(\Delta t-s')}\,ds\,ds'
    \nonumber\\
    &= \int_0^{\Delta t}
    e^{-\mathbf{A}(\Delta t-s)}\,(2\mathbf{B})\,
    e^{-\mathbf{A}^\top(\Delta t-s)}\,ds
    \nonumber\\
    &= \int_0^{\Delta t}
    e^{-\mathbf{A}u}\,(2\mathbf{B})\,
    e^{-\mathbf{A}^\top u}\,du,
    \label{eq:Sdt_calc}
\end{align}
where in the third line we used
$\mathbb{E}[\boldsymbol{\eta}(t)\boldsymbol{\eta}(t')^\top]
= 2\mathbf{B}\,\delta(t-t')$, and in the fourth line we substituted
$u = \Delta t - s$. To show that 
\begin{equation}
    \boldsymbol{\Sigma}_{\Delta t}=\boldsymbol{\Sigma}_\infty - \mathbf{Q}\boldsymbol{\Sigma}_\infty\mathbf{Q}^\top
\end{equation}
we use the continuous Lyapunov equation \eqref{eq:lyapunov_} to write
\begin{align}
    \int_0^{\Delta t}
    e^{-\mathbf{A}u}\,(2\mathbf{B})\,e^{-\mathbf{A}^\top u}\,du
    &= \int_0^{\Delta t}
    e^{-\mathbf{A}u}\!\left(\mathbf{A}\boldsymbol{\Sigma}_\infty
    + \boldsymbol{\Sigma}_\infty\mathbf{A}^\top\right)
    e^{-\mathbf{A}^\top u}\,du
    \nonumber\\
    &= \int_0^{\Delta t}\!\left(
    -\frac{d}{du}e^{-\mathbf{A}u}\right)\boldsymbol{\Sigma}_\infty\,
    e^{-\mathbf{A}^\top u}
    + e^{-\mathbf{A}u}\,\boldsymbol{\Sigma}_\infty
    \left(-\frac{d}{du}e^{-\mathbf{A}^\top u}\right)du
    \nonumber\\
    &= -\int_0^{\Delta t}
    \frac{d}{du}\!\left(e^{-\mathbf{A}u}\,
    \boldsymbol{\Sigma}_\infty\,e^{-\mathbf{A}^\top u}\right)du
    \nonumber\\
    &= \left[-e^{-\mathbf{A}u}\,\boldsymbol{\Sigma}_\infty\,
    e^{-\mathbf{A}^\top u}\right]_0^{\Delta t}
    \nonumber\\
    &= \boldsymbol{\Sigma}_\infty
    - e^{-\mathbf{A}\Delta t}\,\boldsymbol{\Sigma}_\infty\,
    e^{-\mathbf{A}^\top\Delta t}
    \nonumber\\
    &= \boldsymbol{\Sigma}_\infty
    - \mathbf{Q}\,\boldsymbol{\Sigma}_\infty\,\mathbf{Q}^\top.
\end{align}
Equation~\eqref{eq:var1} is exact: the discretely sampled MVOU process
is a Gaussian VAR(1) with no approximation error, and is fully
characterised by the parameter pair
$\boldsymbol{\theta} = (\mathbf{Q}, \boldsymbol{\Sigma}_{\Delta t})$,
which is directly accessible from the data.

Once $\hat{\mathbf{Q}}$ and $\hat{\boldsymbol{\Sigma}}_{\Delta t}$
have been estimated (see Section~\ref{sec:mle}), the continuous-time
parameters are recovered as follows. The drift matrix is obtained via
the matrix logarithm,
\begin{equation}
    \hat{\mathbf{A}} = -\frac{1}{\Delta t}\log\hat{\mathbf{Q}},
    \label{eq:Ahat}
\end{equation}
which is well defined under the stability assumption, since all
eigenvalues of $\mathbf{Q} = e^{-\mathbf{A}\Delta t}$ lie in $(0,1)$.
The stationary covariance $\hat{\boldsymbol{\Sigma}}_\infty$ is then
the solution of the discrete Lyapunov equation
\begin{equation}
    \hat{\boldsymbol{\Sigma}}_\infty =
    \hat{\mathbf{Q}}\,\hat{\boldsymbol{\Sigma}}_\infty\,\hat{\mathbf{Q}}^\top
    + \hat{\boldsymbol{\Sigma}}_{\Delta t},
    \label{eq:disc_lyap}
\end{equation}
and the diffusion matrix is recovered indirectly via the continuous
Lyapunov equation \eqref{eq:lyapunov_},
\begin{equation}
    \hat{\mathbf{B}} = \frac{1}{2}\!\left(
    \hat{\mathbf{A}}\,\hat{\boldsymbol{\Sigma}}_\infty
    + \hat{\boldsymbol{\Sigma}}_\infty\,\hat{\mathbf{A}}^\top
    \right).
    \label{eq:Bhat}
\end{equation}

\section{Random matrix ensembles}\label{sec:ensembles}

\subsection{Gaussian Orthogonal Ensemble}
We focus first on reversible MVOU processes, for which the detailed
balance condition holds, and
as discussed in Section~\ref{sec:model}, one can always perform a
change of variables that
maps the dynamics onto an equivalent system with $\mathbf{B} =
\mathbf{I}$ and $\mathbf{A} = \mathbf{A}^\top$. The simplest ensemble of symmetric random matrices is the Gaussian
Orthogonal Ensemble (GOE), thus we model $\mathbf{A}$ as
\begin{equation}
    \mathbf{A} = \mathbf{I} + \mathbf{W},
    \qquad
    \mathbf{W} \sim \mathrm{GOE}\!\left(0,\,\frac{c^2}{N}\right),
    \label{eq:GOE}
\end{equation}
where the off-diagonal entries of $\mathbf{W}\in\mathbb{R}^{N\times N}$
have variance $c^2/(2N)$ and the diagonal entries have variance
$c^2/N$. In the large-$N$ limit, the
eigenvalue density of $\mathbf{A}$ converges to the Wigner
semicircle law shifted to unit mean,
\begin{equation}
    \rho(\lambda) = \frac{1}{\pi c^2}
    \sqrt{2c^2 - (\lambda-1)^2}\;
    \mathbf{1}_{\{|\lambda - 1| \leq \sqrt{2}\,c\}},
\end{equation}
with support $[1 - \sqrt{2}\,c,\, 1 + \sqrt{2}\,c]$. The system
remains stable as long as the lower spectral edge is positive, which
requires $c < c_{\mathrm{crit}} = 1/\sqrt{2}$.The distance to instability in the large-$N$ limit is therefore
\begin{equation}
    r = 1 - \sqrt{2}\,c.
    \label{eq:r_GOE}
\end{equation}

The fluctuations of the smallest eigenvalue around this value are
governed by the Tracy--Widom distribution. Specifically, for
$\mathbf{A}$ drawn from the GOE ensemble \eqref{eq:GOE}, the
rescaled edge eigenvalue converges in distribution to the
Tracy--Widom$_1$ law~\cite{tw2},
\begin{equation}
    N^{2/3}\,
    \frac{(1-\sqrt{2}\,c)-\lambda_{\min}(\mathbf{A})}
         {(\sqrt{2}\,c)^{1/3}}
    \;\xrightarrow{d}\;
    \mathrm{TW}_1,
    \label{eq:TW_GOE}
\end{equation}
where $\mathrm{TW}_1$ has mean
$\mu_{\mathrm{TW}} \approx -1.21$ and is left-skewed with
skewness $\approx -0.29$. Taking expectations yields
\begin{equation}
    \bar{r}(c,N)
    =
    (1-\sqrt{2}\,c)
    -
    (\sqrt{2}\,c)^{1/3}
    \mu_{\mathrm{TW}}\,N^{-2/3},
    \qquad
    \mu_{\mathrm{TW}}\approx -1.21,
    \label{eq:rbar_}
\end{equation}
so that finite systems are on average slightly more stable than
predicted by the large-$N$ limit. Nevertheless, the finite-width
Tracy--Widom fluctuations imply that a non-zero fraction of
realizations may still cross the stability threshold.

\subsection{Ginibre Ensemble}
For non-reversible MVOU processes, the drift matrix $\mathbf{A}$ is
asymmetric, $\mathbf{A}\mathbf{A}^\top \neq \mathbf{A}^\top\mathbf{A}$,
and its eigenvalues are generically complex. The appropriate ensemble
is the real Ginibre ensemble (GinOE), in which the entries of
$\mathbf{W}\in\mathbb{R}^{N\times N}$ are i.i.d.\ Gaussian,
\begin{equation}
    \mathbf{A} = \mathbf{I} + \mathbf{W},
    \qquad
    W_{ij} \overset{\mathrm{i.i.d.}}{\sim}
    \mathcal{N}\!\left(0,\,\frac{c^2}{N}\right).
    \label{eq:Ginibre}
\end{equation}
In the large-$N$ limit, the eigenvalues of $\mathbf{W}$ are
asymptotically uniform on the disk of radius $c$ in the complex plane
(circular law~\cite{girko}),
\begin{equation}
    \rho(\lambda) = \frac{1}{\pi c^2}\,
    \mathbf{1}_{\{|\lambda| \leq c\}},
    \qquad \lambda\in\mathbb{C}.
    \label{eq:circular_law}
\end{equation}
The eigenvalues of $\mathbf{A} = \mathbf{I} + \mathbf{W}$ are
therefore asymptotically uniform on the disk of radius $c$ centred at
$1$ in the complex plane. The system remains stable as long as no
eigenvalue has non-positive real part, i.e.\ the disk does not
intersect the imaginary axis, which requires $c < c_{\mathrm{crit}} =
1$. In the large-$N$ limit the distance to instability is
\begin{equation}
    r = 1 - c 
    \label{eq:r_Ginibre}
\end{equation}
 These edge fluctuations at finite size are not
governed by the Tracy--Widom$_1$ law but belong to a different
universality class.

\section{Bayesian inference Problem}
\label{sec:mle}

Given the observed trajectory $\mathcal{D} = \{\mathbf{X}_k\}_{k=0}^{M-1}$,
the posterior distribution over parameters is given by Bayes' theorem,
\begin{equation}
    P(\boldsymbol{\theta} \mid \mathcal{D})
    = \frac{P(\mathcal{D} \mid \boldsymbol{\theta})\,P(\boldsymbol{\theta})}
           {P(\mathcal{D})}.
    \label{eq:bayes}
\end{equation}
We adopt a uniform prior $P(\boldsymbol{\theta}) \propto \mathrm{const}$,
encoding the absence of prior information on the parameters. Under this
choice the posterior is proportional to the likelihood,
\begin{equation}
    P(\boldsymbol{\theta} \mid \mathcal{D}) \propto P(\mathcal{D} \mid \boldsymbol{\theta}),
\end{equation}
and inference is entirely determined by the likelihood of the observed
trajectory. The maximum a posteriori (MAP) and maximum likelihood (MLE)
estimators therefore coincide,
\begin{equation}
    \hat{\boldsymbol{\theta}}
    = \operatorname*{arg\,max}_{\boldsymbol{\theta}}\, P(\boldsymbol{\theta} \mid \mathcal{D})
    = \operatorname*{arg\,max}_{\boldsymbol{\theta}}\, P(\mathcal{D} \mid \boldsymbol{\theta}),
\end{equation}
since the two objectives differ only by a constant log-prior and the
$\boldsymbol{\theta}$-independent normalisation $\log P(\mathcal{D})$.
We write $\hat{\boldsymbol{\theta}} = (\hat{\mathbf{Q}},\,
\hat{\boldsymbol{\Sigma}}_{\Delta t})$ for this common solution
throughout.

\medskip

By the Markov property of \eqref{eq:var1}, the likelihood of the full
trajectory factors as
\begin{equation}
    P(\mathcal{D} \mid \boldsymbol{\theta})
    = P(\mathbf{X}_0 \mid \boldsymbol{\theta})
    \prod_{k=0}^{M-2} P(\mathbf{X}_{k+1} \mid \mathbf{X}_k,
    \boldsymbol{\theta}),
    \label{eq:lik_factored}
\end{equation}
where the stationary initial distribution is
\begin{equation}
    P(\mathbf{X}_0 \mid \boldsymbol{\theta})
    = \frac{1}{(2\pi)^{N/2}\det(\boldsymbol{\Sigma}_\infty)^{1/2}}
    \exp\!\left(
    -\frac{1}{2}\mathbf{X}_0^\top \boldsymbol{\Sigma}_\infty^{-1} \mathbf{X}_0
    \right),
    \label{eq:init_dist}
\end{equation}
and each transition density is
\begin{equation}
    P(\mathbf{X}_{k+1} \mid \mathbf{X}_k, \boldsymbol{\theta})
    = \frac{1}{(2\pi)^{N/2}\det(\boldsymbol{\Sigma}_{\Delta t})^{1/2}}
    \exp\!\left(
    -\frac{1}{2}(\mathbf{X}_{k+1} - \mathbf{Q}\mathbf{X}_k)^\top
    \boldsymbol{\Sigma}_{\Delta t}^{-1}
    (\mathbf{X}_{k+1} - \mathbf{Q}\mathbf{X}_k)
    \right).
    \label{eq:trans_dist}
\end{equation}

\medskip

Taking the logarithm of \eqref{eq:lik_factored} and substituting
\eqref{eq:init_dist} and \eqref{eq:trans_dist} gives the full
log-likelihood,
\begin{equation}
\begin{aligned}
    \log P(\mathcal{D} \mid \boldsymbol{\theta})
    &= -\frac{1}{2}\log\det(2\pi\boldsymbol{\Sigma}_\infty)
       -\frac{1}{2}\mathbf{X}_0^\top \boldsymbol{\Sigma}_\infty^{-1} \mathbf{X}_0
       -\frac{M-1}{2}\log\det(2\pi\boldsymbol{\Sigma}_{\Delta t})\\
    &\quad - \frac{1}{2}\sum_{k=0}^{M-2}
    (\mathbf{X}_{k+1} - \mathbf{Q}\mathbf{X}_k)^\top
    \boldsymbol{\Sigma}_{\Delta t}^{-1}
    (\mathbf{X}_{k+1} - \mathbf{Q}\mathbf{X}_k).
\end{aligned}
    \label{eq:loglik_full}
\end{equation}
The first two terms account for the stationary initialisation and
contribute an $O(1)$ correction to the dominant $O(M)$ terms; they
are neglected in what follows.

\subsection{MLE estimators}
Using the trace identity
$\mathbf{x}^\top \mathbf{A}\mathbf{y} = \mathrm{Tr}(\mathbf{A}\,
\mathbf{y}\mathbf{x}^\top)$, the quadratic form in \eqref{eq:loglik_full}
expands as
\begin{equation}
    \sum_{k=0}^{M-2}
    (\mathbf{X}_{k+1} - \mathbf{Q}\mathbf{X}_k)^\top
    \boldsymbol{\Sigma}_{\Delta t}^{-1}
    (\mathbf{X}_{k+1} - \mathbf{Q}\mathbf{X}_k)
    = (M-1)\,\mathrm{Tr}\!\Big[
    \boldsymbol{\Sigma}_{\Delta t}^{-1}
    \big(T_1 - \mathbf{Q}T_2^\top - T_2\mathbf{Q}^\top
    + \mathbf{Q}T_3\mathbf{Q}^\top\big)
    \Big],
    \label{eq:quad_expand}
\end{equation}
where we have defined the three sufficient statistics
\begin{equation}
    T_1 = \frac{1}{M-1}\sum_{k=0}^{M-2}\mathbf{X}_{k+1}\mathbf{X}_{k+1}^\top,
    \qquad
    T_2 = \frac{1}{M-1}\sum_{k=0}^{M-2}\mathbf{X}_{k+1}\mathbf{X}_{k}^\top,
    \qquad
    T_3 = \frac{1}{M-1}\sum_{k=0}^{M-2}\mathbf{X}_k\mathbf{X}_k^\top.
    \label{eq:suff}
\end{equation}
The log-likelihood therefore takes the compact form
\begin{equation}
    \log P(\mathcal{D} \mid \boldsymbol{\theta}) =
    -\frac{M-1}{2}\log\det(2\pi\boldsymbol{\Sigma}_{\Delta t})
    -\frac{M-1}{2}\,\mathrm{Tr}\!\left[
    \boldsymbol{\Sigma}_{\Delta t}^{-1}
    \big(T_1 - \mathbf{Q}T_2^\top - T_2\mathbf{Q}^\top
    + \mathbf{Q}T_3\mathbf{Q}^\top\big)
    \right] + \mathrm{const}.
    \label{eq:loglik_suff}
\end{equation}

\paragraph{Maximisation over \texorpdfstring{$\mathbf{Q}$}{Q}.}
Let $\mathbf{J} = \boldsymbol{\Sigma}_{\Delta t}^{-1} \succ 0$.
Differentiating \eqref{eq:loglik_suff} with respect to $Q_{ab}$,
\begin{equation}
    \frac{\partial\log P(\mathcal{D}\mid\boldsymbol{\theta})}{\partial Q_{ab}}
    = -(M-1)\left[\mathbf{J}
    \big(\mathbf{Q}T_3 - T_2\big)\right]_{ab} = 0,
\end{equation}
which, since $\mathbf{J}\succ 0$, gives
\begin{equation}
    \hat{\mathbf{Q}} = T_2 T_3^{-1}.
    \label{eq:Qhat}
\end{equation}
Since $\boldsymbol{\Sigma}_{\Delta t} \succ 0$, the first term is
non-negative and vanishes exactly at $\hat{\mathbf{Q}} = T_2 T_3^{-1}$.

\paragraph[Maximisation over Sigma]{Maximisation over $\boldsymbol{\Sigma}_{\Delta t}$.}
Substituting \eqref{eq:Qhat} into \eqref{eq:loglik_suff} and
differentiating with respect to $\mathbf{J} =
\boldsymbol{\Sigma}_{\Delta t}^{-1}$, using
$\partial\log\det\mathbf{J}/\partial\mathbf{J} = \mathbf{J}^{-1}$
and $\partial\,\mathrm{Tr}[\mathbf{J}\mathbf{S}]/\partial\mathbf{J} =
\mathbf{S}^\top$,
\begin{equation}
    \frac{\partial\log P(\mathcal{D}\mid\boldsymbol{\theta})}{\partial\mathbf{J}}
    = \frac{M-1}{2}\,\mathbf{J}^{-1}
    - \frac{M-1}{2}
    \big(T_1 - \hat{\mathbf{Q}}T_2^\top - T_2\hat{\mathbf{Q}}^\top
    + \hat{\mathbf{Q}}T_3\hat{\mathbf{Q}}^\top\big)^\top
    = \mathbf{0}.
\end{equation}
Solving for $\boldsymbol{\Sigma}_{\Delta t}$ and substituting
$\hat{\mathbf{Q}} = T_2 T_3^{-1}$,
\begin{equation}
    \hat{\boldsymbol{\Sigma}}_{\Delta t}
    = T_1 - \hat{\mathbf{Q}}T_2^\top - T_2\hat{\mathbf{Q}}^\top
    + \hat{\mathbf{Q}}T_3\hat{\mathbf{Q}}^\top
    = T_1 - T_2 T_3^{-1} T_2^\top,
    \label{eq:Sdthat}
\end{equation}
where the last step uses
$(T_2 T_3^{-1})\,T_3\,(T_2 T_3^{-1})^\top -
T_2\,(T_2 T_3^{-1})^\top - (T_2 T_3^{-1})\,T_2^\top =
-T_2 T_3^{-1} T_2^\top$.

\paragraph{Estimators and consistency.}
The closed-form MLE estimators are
\begin{equation}
    \hat{\mathbf{Q}} = T_2 T_3^{-1},
    \qquad
    \hat{\boldsymbol{\Sigma}}_{\Delta t} = T_1 - T_2 T_3^{-1} T_2^\top,
    \label{eq:estimators}
\end{equation}
which exist as long as $T_3$ is invertible, requiring $q > 1$.
The estimators \eqref{eq:estimators} are consistent: as $M \to \infty$,
$T_2 \to \mathbf{Q}\,\boldsymbol{\Sigma}_\infty$ and $T_3 \to
\boldsymbol{\Sigma}_\infty$ by the ergodic theorem for stationary
processes \cite{singh2018fast}, so $\hat{\mathbf{Q}} \to \mathbf{Q}$ and
$\hat{\mathbf{A}} \to \mathbf{A}$.

\subsection{Hessian}
The precision of parameter estimation is governed by the curvature of
the log-likelihood at its maximum. Substituting the completed-square
identity
\begin{equation}
    T_1 - \mathbf{Q}T_2^\top - T_2\mathbf{Q}^\top + \mathbf{Q}T_3\mathbf{Q}^\top
    = (\mathbf{Q} - T_2 T_3^{-1})\,T_3\,(\mathbf{Q} - T_2 T_3^{-1})^\top
    + \hat{\boldsymbol{\Sigma}}_{\Delta t}
    \label{eq:complete_square}
\end{equation}
into \eqref{eq:loglik_suff}, the log-likelihood reads
\begin{equation}
    \log P(\mathcal{D}\mid\boldsymbol{\theta})
    = \frac{M-1}{2}\log\det\mathbf{J}
    - \frac{M-1}{2}\,\mathrm{Tr}\!\Big[
    \mathbf{J}\Big(
    (\mathbf{Q} - T_2 T_3^{-1})\,T_3\,(\mathbf{Q} - T_2 T_3^{-1})^\top
    + \hat{\boldsymbol{\Sigma}}_{\Delta t}
    \Big)
    \Big] + \mathrm{const}.
    \label{eq:loglik_J}
\end{equation}

\paragraph{Mixed block.}
The mixed second derivative with respect to $J_{cd}$ and $Q_{ab}$ is

\begin{equation}
    \frac{\partial^2 \log P}{\partial J_{cd}\,\partial Q_{ab}}
    = -(M-1)\,(\mathbf{Q}T_3 - T_2)_{ca}\,\delta_{db},
\end{equation}
which is proportional to $\mathbf{Q}T_3 - T_2$ and vanishes at
$\hat{\mathbf{Q}}$. The full Hessian is therefore block diagonal at the
MLE,
\begin{equation}
    \mathbf{H} =
    \begin{pmatrix}
    \mathbf{H}^{(JJ)} & \mathbf{0} \\
    \mathbf{0}        & \mathbf{H}^{(QQ)}
    \end{pmatrix},
    \label{eq:hessian_block}
\end{equation}
so that drift estimation and noise estimation decouple at the optimum.

\paragraph{The $\mathbf{Q}$ block.}
Differentiating the gradient $\partial \log P/\partial \mathbf{Q} =
-(M-1)\,\mathbf{J}\,(\mathbf{Q}T_3 - T_2)$ once more with respect to
$Q_{cd}$, and using $\partial(\mathbf{Q}T_3)_{ca}/\partial Q_{cd} =
(T_3)_{da}$,
\begin{equation}
    -\frac{\partial^2 \log P}{\partial Q_{ab}\,\partial Q_{cd}}
    = (M-1)\,J_{ac}\,(T_3)_{bd}.
\end{equation}
Recognising the index pattern as that of the Kronecker product under the
column-stacking convention $\mathrm{vec}(\mathbf{A}\mathbf{X}\mathbf{B})
= (\mathbf{B}^\top\otimes\mathbf{A})\,\mathrm{vec}(\mathbf{X})$,
\begin{equation}
    \mathbf{H}^{(QQ)} = (M-1)\,(\mathbf{J} \otimes T_3)
    \;
    \label{eq:HQQ}
\end{equation}

\paragraph{The $\mathbf{J}$ block.}
Using $\partial^2 \log\det\mathbf{J}/\partial J_{ab}\,\partial J_{cd}
= -J^{-1}_{bc}\,J^{-1}_{da}$ and the fact that
$\partial^2\,\mathrm{Tr}[\mathbf{J}\mathbf{S}]/\partial J_{ab}\,
\partial J_{cd} = 0$, the Hessian in $\mathbf{J}$ receives contributions
only from the $\log\det$ term,
\begin{equation}
    \left[\mathbf{H}^{(JJ)}\right]_{(ab),(cd)}
    = \frac{M-1}{2}\,J^{-1}_{bc}\,J^{-1}_{da}.
    \label{eq:HJJ_index}
\end{equation}
Rewriting in matrix form requires care with index ordering. With the
convention $[\mathbf{A}\otimes\mathbf{B}]_{(ij),(kl)} = A_{ik}B_{jl}$,
one has $J^{-1}_{bc}J^{-1}_{da} =
[\mathbf{J}^{-\top}\otimes\mathbf{J}^{-1}]_{(ab),(cd)}$, so that
\begin{equation}
    \mathbf{H}^{(JJ)}
    = \frac{M-1}{2}\,\boldsymbol{\Sigma}_{\Delta t}^\top
    \otimes \boldsymbol{\Sigma}_{\Delta t}.
    \label{eq:HJJ}
\end{equation}

\section{Three levels of randomness}
\label{sec:randomness}

The inference problem involves three distinct sources of randomness,
which must be carefully distinguished. For clarity we work throughout
this section in the reversible case with $\bm{B} = \bm{I}$, which is
without loss of generality after the change of variables described in
Section~\ref{sec:model}; the drift matrix is therefore symmetric and
the parameter of interest is the true propagator 
$\bm{Q}_\star = e^{-\bm{A}_\star\Delta t}$, which generated the 
observed trajectory and is to be estimated from it.

\paragraph{Thermal average.}
The first source of randomness is the candidate propagator $\bm{Q}$
itself. To understand its role, it helps to step back and think about
what inference means in this context.

Given a finite observed trajectory $\bm{X}$, we want to estimate the
true propagator $\bm{Q}_\star$. The maximum-likelihood estimator
$\hat{\bm{Q}} = T_2 T_3^{-1}$ provides a point estimate, but it
carries uncertainty: a different trajectory generated by the same
$\bm{Q}_\star$ would yield a different $\hat{\bm{Q}}$. To quantify
this uncertainty, we adopt a Bayesian perspective and treat $\bm{Q}$
itself as a random variable, distributed according to the posterior
\begin{equation}
    p(\bm{Q}\mid\bm{X})
    \propto p(\bm{X}\mid\bm{Q})\,\pi(\bm{Q}),
\end{equation}
where $p(\bm{X}\mid\bm{Q})$ is the likelihood of the observed
trajectory and $\pi(\bm{Q})\propto\mathrm{const}$ is the flat prior.
Since the likelihood is Gaussian in $\bm{Q}$ (Section~\ref{sec:mle}),
the posterior is also Gaussian, concentrated around $\hat{\bm{Q}}$
with width controlled by the curvature of the log-likelihood.

This posterior is precisely a Gibbs measure. Defining the Hamiltonian
as the negative log-likelihood, $\mathcal{H}(\bm{Q}) :=
-\ell(\bm{Q};\bm{X})$, the posterior reads
\begin{equation}
    p_\beta(\bm{Q}\mid\bm{X})
    = \frac{\pi(\bm{Q})\,e^{-\beta\mathcal{H}(\bm{Q})}}{Z_\beta},
    \qquad
    Z_\beta = \int \pi(\bm{Q})\,e^{-\beta\mathcal{H}(\bm{Q})}\,d\bm{Q}.
    \label{eq:gibbs}
\end{equation}
We denote averages with respect to \eqref{eq:gibbs} by
$\langle\,\cdot\,\rangle_\beta$. Under the flat prior
$\pi(\bm{Q})\propto\mathrm{const}$, the prior contributes only a
multiplicative constant and does not shift the mode of the posterior;
the mode therefore coincides with the MLE $\hat{\bm{Q}} = T_2T_3^{-1}$
at any $\beta$. However, the mode is not the full story: at $\beta=1$
the posterior retains finite width around $\hat{\bm{Q}}$, encoding
the estimation uncertainty that a point estimate discards.

The statistical meaning of $\beta$ is best seen by considering three
regimes:
\begin{itemize}

    \item $\beta = 0$: the likelihood is ignored entirely and the
    measure reduces to the prior $\pi(\bm{Q})$, encoding complete
    ignorance about the parameters: the data carry no weight.

    \item $\beta = 1$: the data enter with their correct statistical
    weight, recovering the standard Bayesian posterior $p(\bm{Q}\mid
    \bm{X})$. This is the maximum a posteriori (MAP) regime: the
    distribution has finite width around $\hat{\bm{Q}}$, and its
    covariance
    \begin{equation}
        \mathrm{Cov}_\beta(\mathrm{vec}\,\bm{Q})\big|_{\beta=1}
        = \bm{H}_{QQ}^{-1}
        \label{eq:thermal_cov}
    \end{equation}
    quantifies the full estimation uncertainty. The partition function
    at this value, $Z_1 = p(\bm{X})$, is the marginal likelihood
    (evidence).

    \item $\beta \to \infty$: the Boltzmann weight
    $e^{-\beta\mathcal{H}(\bm{Q})}$ suppresses all configurations
    except the unique minimiser of $\mathcal{H}$. The posterior
    collapses to a Dirac delta centred on $\hat{\bm{Q}}$, the
    covariance \eqref{eq:thermal_cov} vanishes, and the thermal
    average becomes trivial, $\langle f(\bm{Q})\rangle_\beta \to
    f(\hat{\bm{Q}})$ for any continuous $f$. This is the MLE regime:
    estimation uncertainty is lost and inference reduces to a purely
    point estimate.

\end{itemize}

The thermal fluctuations of $\bm{Q}$ around $\hat{\bm{Q}}$ at
$\beta=1$ are therefore a direct measure of estimation uncertainty:
the softer the likelihood curvature, the larger the posterior spread
$\bm{H}_{QQ}^{-1}$, and the harder the inference problem. As $r\to 0$,
the curvature of the likelihood becomes vanishingly small along the
eigendirection of $\bm{H}_{QQ}$ associated with the soft mode,
the direction in parameter space corresponding to the least stable
eigenvalue of $\bm{A}_\star$. The posterior spreads without bound
exclusively along this direction, while remaining well-concentrated
along all others. This directional divergence is the quantitative
expression of critical slowing down of inference: proximity to
instability makes the soft mode unresolvable, while leaving the
remaining modes unaffected.

\paragraph{Quenched disorder: two nested levels.}
The second and third sources of randomness constitute the \emph{quenched
disorder}: unlike the thermal field $\bm{Q}$, they are drawn once and
then held fixed throughout inference. There are two nested levels.

\begin{itemize}

    \item \emph{Structural disorder}: the true drift matrix
    $\bm{A}_\star = \bm{I} + \bm{M}$, with $\bm{M}\sim\mathrm{GOE}(c)$,
    is drawn from the ensemble $P(\bm{A}_\star)$. Since $\bm{Q}_\star =
    e^{-\bm{A}_\star\Delta t}$ is a deterministic function of
    $\bm{A}_\star$ for fixed $\Delta t$, this induces a well-defined
    distribution $P(\bm{Q}_\star)$ on the true propagator. The
    structural disorder therefore fixes $\bm{Q}_\star$ before any data
    are collected, along with the distance to instability
    $r = \min_i\lambda_i(\bm{A}_\star)$, which controls the difficulty
    of the inference problem.

    \item \emph{Observational disorder}: given $\bm{Q}_\star$, a single
    trajectory $\bm{X} = (\bm{X}_0,\dots,\bm{X}_{M-1})$ is drawn from
    the stationary VAR(1) process
    \begin{equation}
        \bm{X}_{k+1} = \bm{Q}_\star\,\bm{X}_k + \bm{\varepsilon}_k,
        \qquad
        \bm{\varepsilon}_k \sim \mathcal{N}(0,\,\bm{\Sigma}_{\Delta t}).
    \end{equation}
    This trajectory determines the sufficient statistics $T_1, T_2,
    T_3$ and hence the location of $\hat{\bm{Q}}$, and is fixed once
    drawn.

\end{itemize}
The full disorder average is therefore nested,
\begin{equation}
    \mathbb{E}[\,\cdot\,]
    = \mathbb{E}_{\bm{Q}_\star}\,
    \mathbb{E}_{\bm{X}\mid\bm{Q}_\star}[\,\cdot\,],
    \label{eq:disorder_avg}
\end{equation}
where the outer expectation averages over $P(\bm{Q}_\star)$ induced
by the GOE ensemble, and the inner expectation averages over
trajectories generated by a given propagator.

\paragraph{Summary.}
The three averages are conceptually and operationally distinct:
\begin{equation}
\underbrace{\langle\,\cdot\,\rangle_\beta}_{\substack{\text{thermal}\\
\bm{Q}\text{ fluctuates}}},
\qquad
\underbrace{\mathbb{E}_{\bm{X}\mid\bm{Q}_\star}[\,\cdot\,]}_{\substack{
\text{observational disorder}\\ \text{trajectory fixed}}},
\qquad
\underbrace{\mathbb{E}_{\bm{Q}_\star}[\,\cdot\,]}_{\substack{
\text{structural disorder}\\ \text{propagator fixed}}}.
\end{equation}
A statement that holds \emph{for a typical trajectory of a typical
system} requires averaging over all three levels. In particular,
it is essential to distinguish whether the disorder average
$\mathbb{E}[\,\cdot\,]$ is taken inside or outside the logarithm of
$Z_\beta$ --- the quenched average $\mathbb{E}[\log Z_\beta]$
describes the typical case, while the annealed average
$\log\mathbb{E}[Z_\beta]$ is dominated by atypical realisations and
is always over-optimistic. 

\subsection{Fisher information and thermal covariance.}
The Fisher information matrix is defined as the average curvature of
the log-likelihood evaluated at the true parameter $\bm{Q}_\star$,
\begin{equation}
    \mathcal{I}(\bm{Q}_\star)
    = -\mathbb{E}_{\bm{X}\mid\bm{Q}_\star}\!\left[
    \frac{\partial^2 \ell(\bm{Q};\bm{X})}
    {\partial\,\mathrm{vec}(\bm{Q})\,\partial\,\mathrm{vec}(\bm{Q})^\top}
    \Bigg|_{\bm{Q}=\bm{Q}_\star}\right].
\end{equation}
Since the log-likelihood is exactly quadratic in $\bm{Q}$
(Section~\ref{sec:mle}), its Hessian is constant --- independent of
the point at which it is evaluated --- and equals $\bm{H}_{QQ}$:
\begin{equation}
    -\frac{\partial^2\ell(\bm{Q};\bm{X})}
    {\partial\,\mathrm{vec}(\bm{Q})\,\partial\,\mathrm{vec}(\bm{Q})^\top}
    = \bm{H}_{QQ} = (M-1)(T_3\otimes\bm{J}).
\end{equation}
Taking the expectation over $\bm{X}$ ,
\begin{equation}
    \mathcal{I}(\bm{Q}_\star)
    = (M-1)\,\mathbb{E}_{\bm{X}\mid\bm{Q}_\star}[T_3]\otimes\bm{J}.
\end{equation}
We compute $\mathbb{E}[T_3]$ directly. By definition,
$T_3 = \frac{1}{M-1}\sum_{k=0}^{M-2}\bm{X}_k\bm{X}_k^\top$, so
\begin{equation}
    \mathbb{E}_{\bm{X}\mid\bm{Q}_\star}[T_3]
    = \frac{1}{M-1}\sum_{k=0}^{M-2}
    \mathbb{E}[\bm{X}_k\bm{X}_k^\top].
\end{equation}
Since the process is stationary with $\bm{X}_k\sim
\mathcal{N}(\bm{0},\bm{\Sigma}_\infty)$ for all $k$
(Section~\ref{sec:model}), we have
$\mathbb{E}[\bm{X}_k\bm{X}_k^\top] = \bm{\Sigma}_\infty$
for every $k$, and therefore
\begin{equation}
    \mathbb{E}_{\bm{X}\mid\bm{Q}_\star}[T_3]
    = \frac{1}{M-1}\sum_{k=0}^{M-2}\bm{\Sigma}_\infty
    = \bm{\Sigma}_\infty.
\end{equation}
Substituting,
\begin{equation}
    \mathcal{I}(\bm{Q}_\star)
    = (M-1)(\bm{\Sigma}_\infty\otimes\bm{J}).
    \label{eq:fisher}
\end{equation}

\begin{remark}
The Fisher information $\mathcal{I}(\bm{Q}_\star) =
(M-1)(\bm{\Sigma}_\infty\otimes\bm{J})$ is conditional on the
structural disorder $\bm{Q}_\star$: it characterises the difficulty
of inference for a given system. A fully disorder-averaged quantity
would require computing $\mathbb{E}_{\bm{Q}_\star}[\mathcal{I}
(\bm{Q}_\star)]$, which involves averaging $\bm{\Sigma}_\infty$ and
$\bm{J} = \bm{\Sigma}_{\Delta t}^{-1}$ over the GOE. 
\end{remark}


\section{Covariance of the estimator and the inflation factor}
\label{sec:kappa}

\paragraph{Exact conditional covariance.}
We derive the sampling covariance of $\hat{\bm{Q}}$ conditional on
the observed regressor path $\{\bm{X}_k\}$. Substituting the VAR(1)
representation \eqref{eq:var1} into the definition of $T_2$,
\begin{equation}
    T_2 = \bm{Q}_\star T_3 + \bm{E},
    \qquad
    \bm{E} := \frac{1}{M-1}\sum_{k=0}^{M-2}
    \bm{\varepsilon}_k\bm{X}_k^\top,
\end{equation}
so that the estimation error satisfies
\begin{equation}
    \hat{\bm{Q}} - \bm{Q}_\star
    = \bm{E}\,T_3^{-1}.
    \label{eq:err}
\end{equation}
Conditional on $\{\bm{X}_k\}$, the innovation matrix $\bm{E}$ is a
sum of independent terms, each linear in $\bm{\varepsilon}_k \sim
\mathcal{N}(\bm{0}, \bm{\Sigma}_{\Delta t})$, with
$\mathbb{E}[\bm{E}\mid\bm{X}] = \bm{0}$. Using the vec identity
$\mathrm{vec}(\bm{\varepsilon}_k\bm{X}_k^\top) =
(\bm{X}_k\otimes\bm{I}_N)\bm{\varepsilon}_k$ and the independence of
the innovations across $k$,
\begin{equation}
    \mathrm{Cov}(\mathrm{vec}\,\bm{E}\mid\bm{X})
    = \frac{1}{(M-1)^2}\sum_{k=0}^{M-2}
    (\bm{X}_k\otimes\bm{I}_N)\,\bm{\Sigma}_{\Delta t}\,
    (\bm{X}_k\otimes\bm{I}_N)^\top
    = \frac{1}{M-1}(T_3\otimes\bm{\Sigma}_{\Delta t}),
\end{equation}
where we used the mixed-product property
$(\bm{A}\otimes\bm{B})(\bm{C}\otimes\bm{D}) = \bm{AC}\otimes\bm{BD}$
and the definition of $T_3$. Since
$\mathrm{vec}(\hat{\bm{Q}}-\bm{Q}_\star) =
(T_3^{-1}\otimes\bm{I}_N)\,\mathrm{vec}\,\bm{E}$ from \eqref{eq:err},
\begin{equation}
    \mathrm{Cov}(\mathrm{vec}\,\hat{\bm{Q}}\mid\bm{X})
    = \frac{1}{M-1}(T_3^{-1}\otimes\bm{\Sigma}_{\Delta t})
    = \bm{H}_{QQ}^{-1}.
    \label{eq:bridge}
\end{equation}
This identity is exact and holds conditional on the regressor path.
Note that although the regressors $\bm{X}_k$ are strongly
autocorrelated, the numerator $\bm{E}$ carries no autocorrelation
penalty: each innovation $\bm{\varepsilon}_k$ is independent of
$\bm{X}_k$ and of the entire past, so all temporal cross-terms
vanish. Temporal correlations enter only through the denominator
$T_3^{-1}$, via how noisily the sample covariance is estimated from
correlated data.

The conditional covariance \eqref{eq:bridge} is exact but depends on
the specific observed trajectory through the random sufficient
statistic $T_3$. To obtain the unconditional covariance, averaged
over all possible trajectories generated by $\bm{Q}_\star$, one must
compute $\mathbb{E}_{\bm{X}\mid\bm{Q}_\star}[\bm{H}_{QQ}^{-1}]$. It
is at this step that an inflation factor emerges, because
$\mathbb{E}[T_3^{-1}] \neq (\mathbb{E}[T_3])^{-1} =
\bm{\Sigma}_\infty^{-1}$.

\paragraph{The origin of $\kappa$: Jensen's inequality and the
quenched versus annealed average.}
The discrepancy between $\mathbb{E}[T_3^{-1}]$ and
$(\mathbb{E}[T_3])^{-1}$ has two equivalent readings.

From the probabilistic point of view, matrix inversion is a convex
operation, and Jensen's inequality gives directly
\begin{equation}
    \mathbb{E}[T_3^{-1}] \geq
    \left(\mathbb{E}[T_3]\right)^{-1} = \bm{\Sigma}_\infty^{-1}.
    \label{eq:jensen}
\end{equation}
Downward fluctuations of $T_3$ inflate $T_3^{-1}$ more than upward
fluctuations deflate it, so the mean of the inverse strictly exceeds
the inverse of the mean.

From the statistical mechanics point of view, this is the quenched
versus annealed distinction of Section~\ref{sec:randomness}. The
annealed estimate of the precision plugs in the mean
$\mathbb{E}[T_3] = \bm{\Sigma}_\infty$ before inverting,
\begin{equation}
    \text{annealed:}\quad
    \left(\mathbb{E}_{\bm{X}\mid\bm{Q}_\star}
    [\bm{H}_{QQ}]\right)^{-1}
    = \frac{1}{M-1}
    (\bm{\Sigma}_\infty^{-1}\otimes\bm{\Sigma}_{\Delta t}).
    \label{eq:annealed}
\end{equation}
This corresponds to an oracle that knows $\bm{\Sigma}_\infty$ exactly
and substitutes it before inverting --- it is over-optimistic because
it ignores the fluctuations of $T_3$ around its mean. The quenched
average instead correctly accounts for those fluctuations by averaging
the inverse after it has been computed from the data,
\begin{equation}
    \text{quenched:}\quad
    \mathbb{E}_{\bm{X}\mid\bm{Q}_\star}[\bm{H}_{QQ}^{-1}]
    = \frac{1}{M-1}\,\mathbb{E}[T_3^{-1}]
    \otimes\bm{\Sigma}_{\Delta t}.
    \label{eq:quenched}
\end{equation}
The two readings, Jensen's inequality and the quenched/annealed
distinction, are two languages for the same phenomenon. The ratio
between the quenched and annealed precisions defines the inflation
factor,
\begin{equation}
    \kappa
    := \frac{\left\|\mathbb{E}_{\bm{X}\mid\bm{Q}_\star}
    [\bm{H}_{QQ}^{-1}]\right\|_F}
    {\left\|\left(\mathbb{E}_{\bm{X}\mid\bm{Q}_\star}
    [\bm{H}_{QQ}]\right)^{-1}\right\|_F}
    \geq 1,
    \label{eq:kappa_ratio}
\end{equation}

where $\|\cdot\|_F$ denotes the Frobenius norm. In our particular case the two matrices
are proportional, so we obtain \eqref{eq:72}.

\paragraph{The inverse-Wishart inflation factor $\kappa_W$.}
To compute $\kappa$ explicitly, we begin with the scalar case as
intuition. Suppose one estimates a variance $\sigma^2$ from $n$
i.i.d.\ Gaussian observations via the sample variance
$s^2 = \frac{1}{n}\sum_k x_k^2$. The estimator $s^2$ is unbiased,
$\mathbb{E}[s^2] = \sigma^2$, but the quantity we need is its
reciprocal $1/s^2$. Since $x\mapsto 1/x$ is convex, Jensen's
inequality gives immediately
\begin{equation}
    \mathbb{E}\!\left[\frac{1}{s^2}\right]
    \geq \frac{1}{\mathbb{E}[s^2]} = \frac{1}{\sigma^2}.
\end{equation}
The inequality is strict, and the exact value follows from
$ns^2/\sigma^2\sim\chi^2_n$ and $\mathbb{E}[1/\chi^2_n] =
1/(n-2)$ for $n>2$,
\begin{equation}
    \mathbb{E}\!\left[\frac{1}{s^2}\right]
    = \frac{1}{\sigma^2}\cdot\frac{n}{n-2}.
    \label{eq:scalar_inflate}
\end{equation}
The precision is therefore biased upward even though the variance
estimate is unbiased: downward fluctuations of $s^2$ inflate $1/s^2$
more than upward fluctuations deflate it, and Jensen's inequality
captures exactly this asymmetry.

The matrix analogue of \eqref{eq:scalar_inflate} is the
inverse-Wishart moment. Under the assumption that the regressors
$\{\bm{X}_k\}$ are i.i.d.\ $\mathcal{N}(\bm{0},\bm{\Sigma}_\infty)$,
the sample covariance satisfies
\begin{equation}
    (M-1)T_3 \sim \mathcal{W}_N(M-1,\bm{\Sigma}_\infty),
\end{equation}
the Wishart distribution with $M-1$ degrees of freedom and scale
matrix $\bm{\Sigma}_\infty$. Jensen's inequality extends to the
matrix setting: since matrix inversion is a convex operation,
\begin{equation}
    \mathbb{E}[T_3^{-1}] \geq \left(\mathbb{E}[T_3]\right)^{-1}
    = \bm{\Sigma}_\infty^{-1},
\end{equation}
and the exact value follows from the inverse moment of the Wishart
distribution: for $M-1 > N+1$,
\begin{equation}
    \mathbb{E}[T_3^{-1}]
    = \frac{M-1}{M-N-2}\,\bm{\Sigma}_\infty^{-1}.
    \label{eq:72}
\end{equation}
Defining the inflation factor
\begin{equation}
    \kappa_W := \frac{M-1}{M-N-2}
    \xrightarrow{\;q\to\infty\;} \frac{q}{q-1},
    \label{eq:kW}
\end{equation}
where $q = (M-1)/N$ is the observation ratio, we can write
$\mathbb{E}[T_3^{-1}] = \kappa_W\,\bm{\Sigma}_\infty^{-1}$.

Substituting into \eqref{eq:bridge} and averaging over the
trajectory,
\begin{equation}
    \mathrm{Cov}(\mathrm{vec}\,\hat{\bm{Q}})
    = \frac{\kappa_W}{M-1}
    (\bm{\Sigma}_\infty^{-1}\otimes\bm{\Sigma}_{\Delta t}).
    \label{eq:cov_kW_}
\end{equation}
The unconditional covariance of the estimator is therefore inflated
by $\kappa_W \geq 1$ relative to the oracle bound obtained by
replacing $T_3$ with its mean $\bm{\Sigma}_\infty$.

All results in this section are conditional on the structural disorder
$\bm{Q}_\star$: the averages $\mathbb{E}_{\bm{X}\mid\bm{Q}_\star}[\,\cdot\,]$
are taken over trajectories generated by a fixed system, consistently
with the nested disorder structure of Section~\ref{sec:randomness}.

\subsection{Spectral decomposition of the estimator covariance and 
propagation to the eigenvalues of $\hat{\bm{A}}$}
\label{sec:vardist}

\paragraph{Spectrum of the unconditional covariance in the reversible case.}
In the reversible case $\bm{B}=\bm{I}$ and $\bm{A}=\bm{A}^\top$,
the drift matrix admits the spectral decomposition
$\bm{A} = \bm{U}\,\mathrm{diag}(\lambda_1,\dots,\lambda_N)\,\bm{U}^\top$
with $\bm{U}$ orthogonal. All matrices entering the problem are
simultaneously diagonalised by $\bm{U}$, with eigenvalues
\begin{equation}
    \mu_a(\bm{\Sigma}_\infty) = \frac{1}{\lambda_a},
    \qquad
    \mu_a(\bm{\Sigma}_{\Delta t})
    = \frac{1-e^{-2\lambda_a\Delta t}}{\lambda_a},
    \qquad
    \mu_a(\bm{\Sigma}_\infty^{-1}) = \lambda_a.
    \label{eq:spectra}
\end{equation}
The unconditional covariance derived in Section~\ref{sec:kappa} is
\begin{equation}
    \mathrm{Cov}(\mathrm{vec}\,\hat{\bm{Q}})
    = \frac{\kappa_W}{M-1}
    (\bm{\Sigma}_\infty^{-1}\otimes\bm{\Sigma}_{\Delta t}).
    \label{eq:cov_kW2_}
\end{equation}
Since $\bm{\Sigma}_\infty^{-1}$ and $\bm{\Sigma}_{\Delta t}$ are both
diagonal in the eigenbasis of $\bm{A}$, the eigenvalues of the
Kronecker product $\bm{\Sigma}_\infty^{-1}\otimes\bm{\Sigma}_{\Delta t}$
are the pairwise products of their eigenvalues:
\begin{equation}
    \mu_{ab}(\bm{\Sigma}_\infty^{-1}\otimes\bm{\Sigma}_{\Delta t})
    = \mu_a(\bm{\Sigma}_\infty^{-1})\cdot\mu_b(\bm{\Sigma}_{\Delta t})
    = \lambda_a\cdot\frac{1-e^{-2\lambda_b\Delta t}}{\lambda_b},
    \qquad a,b=1,\dots,N.
\end{equation}
Substituting into \eqref{eq:cov_kW2_},
\begin{equation}
    \mu_{ab}\!\left(\mathrm{Cov}(\mathrm{vec}\,\hat{\bm{Q}})\right)
    = \frac{\kappa_W}{M-1}\cdot
    \frac{\lambda_a\,(1-e^{-2\lambda_b\Delta t})}{\lambda_b}.
    \label{eq:cov_spec}
\end{equation}

\paragraph{Variance of the projected propagator.}
The diagonal entries of $\mathrm{Cov}(\mathrm{vec}\,\hat{\bm{Q}})$
give the variance of the individual matrix elements of $\hat{\bm{Q}}$
in the eigenbasis of $\bm{A}$. In the reversible case, the
eigenvectors $\{\bm{u}_b\}$ of $\bm{A}$ are known exactly and shared
by all matrices in the problem, so the projected propagator estimate
along mode $b$,
\begin{equation}
    \hat{q}_b = \bm{u}_b^\top\hat{\bm{Q}}\,\bm{u}_b,
\end{equation}
is well defined and its variance is read off from the diagonal entry
$a=b$ of \eqref{eq:cov_spec},
\begin{equation}
    \mathrm{Var}(\hat{q}_b)
    = \frac{\kappa_W}{M-1}\cdot
    \frac{\lambda_b\,(1-e^{-2\lambda_b\Delta t})}{\lambda_b}
    = \frac{\kappa_W\,(1-e^{-2\lambda_b\Delta t})}{M-1}
    = \frac{\kappa_W\,(1-q_b^2)}{M-1},
\end{equation}
where $q_b = e^{-\lambda_b\Delta t}$.

\paragraph{Variance of the eigenvalues of $\hat{\bm{A}}$.}
The estimated drift matrix is $\hat{\bm{A}} = -\frac{1}{\Delta t}
\log\hat{\bm{Q}}$, so the estimated eigenvalue of mode $b$ is
\begin{equation}
    \hat{\lambda}_b = -\frac{1}{\Delta t}\log\hat{q}_b.
\end{equation}
We propagate the variance of $\hat{q}_b$ to $\hat{\lambda}_b$ via
the delta method. The function $f(q) = -\frac{1}{\Delta t}\log q$
is smooth in a neighbourhood of the true value $q_b$, and the
first-order approximation is accurate to $O((M-1)^{-1})$ for large
$M$. Evaluating the derivative at the true value $q_b$,
\begin{equation}
    \frac{df}{dq}\bigg|_{q=q_b}
    = -\frac{1}{\Delta t\,q_b},
\end{equation}
the delta method gives
\begin{equation}
    \mathrm{Var}(\hat{\lambda}_b)
    \approx \left(\frac{df}{dq}\bigg|_{q=q_b}\right)^2
    \mathrm{Var}(\hat{q}_b)
    = \frac{1}{\Delta t^2\,q_b^2}\cdot
    \frac{\kappa_W\,(1-q_b^2)}{M-1}.
\end{equation}
Substituting $q_b = e^{-\lambda_b\Delta t}$,
\begin{equation}
    \mathrm{Var}(\hat{\lambda}_b)
    = \frac{\kappa_W}{M-1}\cdot
    \frac{1-e^{-2\lambda_b\Delta t}}{\Delta t^2\,e^{-2\lambda_b\Delta t}}
    = \frac{\kappa_W}{M-1}\cdot
    \frac{e^{2\lambda_b\Delta t}-1}{\Delta t^2}.
\end{equation}
This is the exact expression for the variance of the estimated
eigenvalue of mode $b$, valid for any $\lambda_b\Delta t > 0$.

\section{Effective sample size and the compound-Wishart correction}
\label{sec:effss}

The inverse-Wishart factor $\kappa_W$ derived in Section~\ref{sec:kappa}
assumed that the $M-1$ vectors forming $T_3$ are i.i.d. They are not:
they are consecutive snapshots of a process with memory. In the
reversible case the process decouples into $N$ independent scalar
AR(1) modes, and the relevant autocorrelation is not that of the
process itself but of its \emph{square}, since $T_3$ is a sum of
outer products $\bm{X}_k\bm{X}_k^\top$.

\paragraph{Autocorrelation time of the variance statistic.}
For mode $a$, the level autocorrelation at lag $h$ is
\begin{equation}
    \mathrm{Corr}(x_{a,t},\, x_{a,t+h}) = q_a^{|h|}.
    \label{eq:acf_level}
\end{equation}
By Isserlis' theorem, since $(x_{a,t}, x_{a,t+h})$ is jointly
zero-mean Gaussian,
\begin{equation}
    \mathbb{E}[x_{a,t}^2\, x_{a,t+h}^2]
    = \mathbb{E}[x_{a,t}^2]\,\mathbb{E}[x_{a,t+h}^2]
    + 2\,\mathbb{E}[x_{a,t}\, x_{a,t+h}]^2,
\end{equation}
so that
\begin{equation}
    \mathrm{Cov}(x_{a,t}^2,\, x_{a,t+h}^2)
    = 2\,\mathrm{Cov}(x_{a,t},\, x_{a,t+h})^2
    = 2s_a^2\,q_a^{2|h|},
\end{equation}
where $s_a = 1/\lambda_a$ is the stationary variance of mode $a$.
Since $\mathrm{Var}(x_a^2) = 2s_a^2$, the autocorrelation of the
squared process is
\begin{equation}
    \mathrm{Corr}(x_{a,t}^2,\, x_{a,t+h}^2) = q_a^{2|h|}.
    \label{eq:acf_square}
\end{equation}
The integrated autocorrelation time of $x_a^2$ is therefore
\begin{equation}
    \tau_a
    = \sum_{h=-\infty}^{+\infty} q_a^{2|h|}
    = \frac{1 + q_a^2}{1 - q_a^2}
    = \frac{1 + e^{-2\lambda_a\Delta t}}{1 - e^{-2\lambda_a\Delta t}},
    \label{eq:tau_a}
\end{equation}
where the geometric series converges since $|q_a| < 1$ for all
stable modes. As $\lambda_a\to\infty$ (stiff modes),
$e^{-2\lambda_a\Delta t}\to 0$ and $\tau_a\to 1$, recovering
the i.i.d.\ limit; as $\lambda_a\to 0$ (soft mode, criticality),
$e^{-2\lambda_a\Delta t}\to 1$ and $\tau_a\to\infty$. Hence
$\tau_a\in[1,\infty)$, with the soft mode always having the
largest autocorrelation time $\tau_r = \tau_{\max}$.

For $M$ large, the effective number of independent observations
along mode $a$ is
\begin{equation}
    M_{\mathrm{eff}}^{(a)}
    = \frac{M}{\tau_a},
    \label{eq:Meff_a}
\end{equation}
where the approximation $M-1\approx M$ introduces an error of
$O(1/M)$, negligible in the large-$M$ limit. Since
$\tau_a \in [1, \infty)$, we have $M_{\mathrm{eff}}^{(a)} \in
(0, M]$.

In practice, $\lambda_{\min}(\mathbf{A})$ is extracted from the
full estimated matrix $\hat{\mathbf{A}}$, so the relevant
effective sample size is not the mode-specific
$M_{\mathrm{eff}}^{(a)}$ but a single global quantity. Since
slow modes contribute disproportionately to the reduction of
the inferential budget, we use the variance-to-mean ratio of
the autocorrelation times,
\begin{equation}
    M_{\rm eff}
    = \frac{M}{\langle\tau\rangle_{\rm w}},
    \qquad
    \langle\tau\rangle_{\rm w}
    = \frac{\sum_a \tau_a^2}{\sum_a \tau_a},
\end{equation}
which gives more weight to the slowest modes. A lower bound on
$M_{\rm eff}$ is given by the soft mode alone,
\begin{equation}
    M_{\rm eff}^{\rm min}
    = \frac{M}{\tau_r}
    \approx Mr\Delta t,
\end{equation}
since $\tau_r = \tau_{\max} \geq \langle\tau\rangle_{\rm w}$.

\subsection{Correction to $\texorpdfstring{\kappa_W}{kappaW}$: the compound-Wishart factor.}

Since the observations are temporally correlated, the standard
Wishart approximation based on $M-1$ independent observations
is no longer strictly valid. A naive but effective correction is
to replace the nominal sample size with the effective sample size
$M_{\rm eff}$ introduced in the previous section. In this way,
the reduction in the number of statistically independent
observations induced by temporal correlations is incorporated
into the theory.

The corresponding global compound-Wishart factor is then

\begin{equation}
    \kappa_{\rm cW}
    = \frac{M_{\rm eff} - 1}
           {M_{\rm eff} - N - 2},
    \label{eq:kcW_global}
\end{equation}
defined for $M_{\rm eff} > N+2$; it diverges as $M_{\rm eff} \to
(N+2)^+$, the minimum number of effective observations required
to estimate the mean, covariance, and cross-covariance of an
$N$-dimensional system. This replaces $\kappa_W$ in
\eqref{eq:cov_kW_} to give a corrected unconditional covariance
\begin{equation}
    \mathrm{Cov}(\mathrm{vec}\,\hat{\bm{Q}})
    \approx \frac{\kappa_{\rm cW}}{M-1}
    (\bm{\Sigma}_\infty^{-1}\otimes\bm{\Sigma}_{\Delta t}).
    \label{eq:cov_kcW}
\end{equation}
An upper bound on $\kappa_{\rm cW}$ is obtained by replacing
$\langle\tau\rangle_{\rm w}$ with the autocorrelation time of
the soft mode alone, $\tau_r = \tau_{\max}$, giving
$M_{\rm eff}^{\rm min} = M/\tau_r \approx Mr\Delta t$ and
\begin{equation}
    \kappa_{\rm cW}^{\rm max}
    = \frac{M/\tau_r - 1}
           {M/\tau_r - N - 2},
\end{equation}
defined for $M/\tau_r > N+2$. Since $\tau_r \geq
\langle\tau\rangle_{\rm w}$, we have $M_{\rm eff}^{\rm min} \leq
M_{\rm eff}$ and therefore $\kappa_{\rm cW}^{\rm max} \geq
\kappa_{\rm cW}$. 
Both bounds recover $\kappa_W$ as $\Delta t \to \infty$, when
consecutive observations become asymptotically independent, and
both diverge when the effective sample size approaches the
breakdown threshold $N+2$. Deriving the exact form of the
compound-Wishart inflation factor for correlated observations
is a non-trivial problem that lies beyond the scope of the
present work.

\section{Derivation of the Frobenius error decomposition}

We derive the decomposition of the squared Frobenius error 
$\|\hat{\mathbf{A}} - \mathbf{A}\|_F^2$ in terms of eigenvalue 
errors, under the assumption of approximate eigenvector alignment.  Let 
$\hat{\mathbf{A}} = U\hat{D}U^\top$ and $\mathbf{A} = VDV^\top$ 
be the eigendecompositions of the estimated and true drift 
matrices, where $\hat{D}$ and $D$ are diagonal matrices of 
eigenvalues and $U$, $V$ are orthogonal matrices of eigenvectors. 
Define the eigenvector overlap matrix $O = U^\top V$.

The squared Frobenius error is
\begin{equation}
\epsilon_A
= \frac{1}{N}\|\hat{\mathbf{A}} - \mathbf{A}\|_F^2
= \frac{1}{N}\mathrm{Tr}\!\left[(\hat{\mathbf{A}} - \mathbf{A})^2\right]
= \frac{1}{N}\left(\mathrm{Tr}(\hat{D}^2) + \mathrm{Tr}(D^2) 
- 2\,\mathrm{Tr}(\hat{D}\,ODO^\top)\right).
\end{equation}
The cross term expands as
\begin{align}
\mathrm{Tr}(\hat{D}\,ODO^\top)
&= \sum_\alpha (\hat{D}\,ODO^\top)_{\alpha\alpha} \notag\\
&= \sum_{\alpha\beta} (\hat{D}\,OD)_{\alpha\beta} O^\top_{\beta\alpha} \notag\\
&= \sum_{\alpha\beta\kappa} (\hat{D}\,O)_{\alpha\kappa} D_{\kappa\beta} O^\top_{\beta\alpha} \notag\\
&= \sum_{\alpha\beta\kappa j} \hat{D}_{\alpha j}\, O_{j\kappa}\, D_{\kappa\beta}\, O^\top_{\beta\alpha} \notag\\
&= \sum_{\alpha\beta\kappa j} \hat{D}_\alpha\,\delta_{\alpha j}\, O_{j\kappa}\, D_\kappa\,\delta_{\kappa\beta}\, O_{\alpha\beta} \notag\\
&= \sum_{\alpha\beta} \hat{D}_\alpha\, D_\beta\, O_{\alpha\beta}\, O_{\alpha\beta} \notag\\
&= \sum_{\alpha\beta} \hat{D}_\alpha\, D_\beta\, |O_{\alpha\beta}|^2,
\end{align}
where we used the diagonality of $\hat{D}$ and $D$ to set 
$\hat{D}_{\alpha j} = \hat{D}_\alpha\delta_{\alpha j}$ and 
$D_{\kappa\beta} = D_\kappa\delta_{\kappa\beta}$. Substituting 
back,
\begin{equation}
\epsilon_A
= \frac{1}{N}\left(
\sum_\alpha \hat{D}_\alpha^2
+ \sum_\alpha D_\alpha^2
- 2\sum_{\alpha\beta} \hat{D}_\alpha D_\beta\, |O_{\alpha\beta}|^2
\right).
\end{equation}
Here $O_{\alpha\beta}^2$ measures the alignment between 
eigenvectors of $\hat{\mathbf{A}}$ and $\mathbf{A}$: 
$|O_{\alpha\beta}|^2 = 1$ for perfectly aligned modes and 
$|O_{\alpha\beta}|^2 = 0$ for orthogonal ones. In the reversible 
case, $\mathbf{A}$ and $\hat{\mathbf{A}}$ are both symmetric and 
to leading order in the estimation error the eigenvectors are 
approximately aligned, $O \approx I$, so that 
$|O_{\alpha\beta}|^2 \approx \delta_{\alpha\beta}$. 
Substituting this approximation yields
\begin{equation}
\epsilon_A
\approx \frac{1}{N}\sum_\alpha (\hat{D}_\alpha - D_\alpha)^2
= \frac{1}{N}\sum_\alpha (\hat{\lambda}_\alpha - \lambda_\alpha)^2.
\end{equation}
The approximation becomes exact 
when $U = V$, i.e.\ when the estimated and true eigenvectors 
coincide.


\newpage

\section{Numerical experiments on real data}
\label{sec:data}

We validate the theoretical predictions of on three
empirical datasets of heterogeneous origin.

\paragraph{Intracranial EEG.}
The first dataset comprises intracranial EEG recordings from 16
epileptic patients; here we analyse a representative single-patient
recording. It contains one seizure and is structured as a preictal
segment (3 min), an ictal segment of variable duration, and a
postictal segment (3 min), sampled at $f_s = 512$\,Hz across
$M = 47$ electrodes ($T = 248{,}320$ samples), giving
$\Delta t = 1/512 \approx 0.002$\,s. Each channel is centred by
subtracting its temporal mean; the resulting matrix is then
rescaled by a single global standard deviation.

\paragraph{Financial returns.}
The second dataset consists of intraday log-returns of $M = 47$
large-cap US equities over the period 2--13 March 2020, spanning
the onset of the COVID-19 market dislocation ($T = 234{,}000$
samples at $\Delta t = 1$\,s). Log-returns are centred per asset
and rescaled by a single global standard deviation; absolute values
are then taken as volatility proxies to capture volatility
clustering. The analysis uses non-overlapping within-day windows to
avoid overnight gaps and artificial serial dependence.

\paragraph{Coastal microbiome.}
The third dataset is the marine bacterial abundance time series, consisting of relative abundances of
$\sim$28{,}600 bacterial OTUs measured three times daily over 88
consecutive days. We retain OTUs present in at least 90\% of
samples, then select the top 47 by mean abundance, yielding a
$264 \times 47$ complete abundance matrix ($\Delta t = 1/3$\,day
$\approx 8$\,h). Zero values remaining after filtering are
replaced by half the minimum nonzero value of the corresponding
species. Abundances are log-transformed, centred per species, and
rescaled by a single global standard deviation.

\noindent
Despite their different origins---neural dynamics, financial
volatility, and ecological community fluctuations---all three
systems are modelled as multivariate Ornstein--Uhlenbeck processes
and analysed with the same estimation pipeline.

\subsection{Estimation pipeline}
\label{sec:pipeline}

\paragraph{Sliding windows.}
For the EEG and financial datasets, estimation is performed on
sliding windows of $T = qN$ samples with a step size of $T/10$,
where $q$ is chosen per figure to balance temporal resolution
against estimation accuracy; the value of $q$ used is reported
in each figure caption. For the financial data, windows are
restricted to within-day segments to avoid overnight gaps and
artificial serial dependence. For the plankton dataset, the
limited number of time points ($T = 264$) does not permit
meaningful windowing, and a single global estimate is computed
on the full time series.

\paragraph{Identifiability check.}
Before running the estimator on each dataset, we verify that the
inference problem is identifiable. The necessary condition is
$q = T/N > 1$, where $T$ is the number of time points in the
window and $N$ is the number of variable. In all cases $\hat{\bm{T}}_3$ is
positive definite and well-conditioned.

\paragraph{Bayesian-I estimator.}
We estimate the drift matrix $\bm{A}$, the propagator
$\bm{Q} = e^{-\bm{A}\Delta t}$, the incremental noise covariance
$\bm{\Sigma}_{\Delta t}$, the stationary covariance
$\bm{\Sigma}_\infty$, and the diffusion matrix
$\bm{B} = \frac{1}{2}(\bm{A}\bm{\Sigma}_\infty +
\bm{\Sigma}_\infty \bm{A}^\top)$ using the Bayesian-I method at
lag $\ell = 1$. Given a data matrix $\bm{X} \in \mathbb{R}^{N
\times T}$, the estimator forms the three empirical moment
matrices
\begin{equation}
    \hat{\bm{T}}_1 = \frac{1}{T-1}\sum_{t=2}^{T}
    \bm{x}_t \bm{x}_t^\top, \quad
    \hat{\bm{T}}_2 = \frac{1}{T-1}\sum_{t=2}^{T}
    \bm{x}_t \bm{x}_{t-1}^\top, \quad
    \hat{\bm{T}}_3 = \frac{1}{T-1}\sum_{t=2}^{T}
    \bm{x}_{t-1} \bm{x}_{t-1}^\top,
\end{equation}
and sets
\begin{equation}
    \hat{\bm{Q}} = \hat{\bm{T}}_2 \hat{\bm{T}}_3^{-1}, \qquad
    \hat{\bm{\Sigma}}_{\Delta t} =
    \hat{\bm{T}}_1 - \hat{\bm{T}}_2
    \hat{\bm{T}}_3^{-1} \hat{\bm{T}}_2^\top, \qquad
    \hat{\bm{A}} = -\frac{1}{\Delta t}\log \hat{\bm{Q}},
\end{equation}
where $\log$ denotes the principal matrix logarithm. The
stationary covariance is obtained by solving the discrete Lyapunov
equation $\hat{\bm{\Sigma}}_\infty = \hat{\bm{Q}}\,
\hat{\bm{\Sigma}}_\infty\,\hat{\bm{Q}}^\top +
\hat{\bm{\Sigma}}_{\Delta t}$.

\paragraph{Eigenvalue spectra and the soft mode.}
For each estimated drift matrix $\hat{\bm{A}}$ we compute the full
eigenvalue spectrum and track the minimum real eigenvalue
$\lambda_{\min}(\hat{\bm{A}})$ across windows. The quantity of interest
is the dimensionless product $r\Delta t =
\lambda_{\min}(\hat{\bm{A}})\,\Delta t$, which measures proximity
to criticality in units of the sampling interval. For the EEG
dataset, $r\Delta t$ is computed for each sliding window and
plotted as a function of time, with the seizure onset marked as a
reference; a sign change of $r\Delta t$ across the seizure
boundary would indicate a transition through criticality. For the
financial dataset, the same quantity is computed within each
trading day and plotted as a function of intraday time, with all
ten days of the COVID-19 dislocation period superimposed. For the
plankton dataset, a single value of $r\Delta t$ is obtained from
the global estimate; we report its sign as an indicator of whether
the bacterial community operates in a stable or near-critical
regime.

\begin{figure}
    \centering
    \includegraphics[width=0.95\columnwidth]{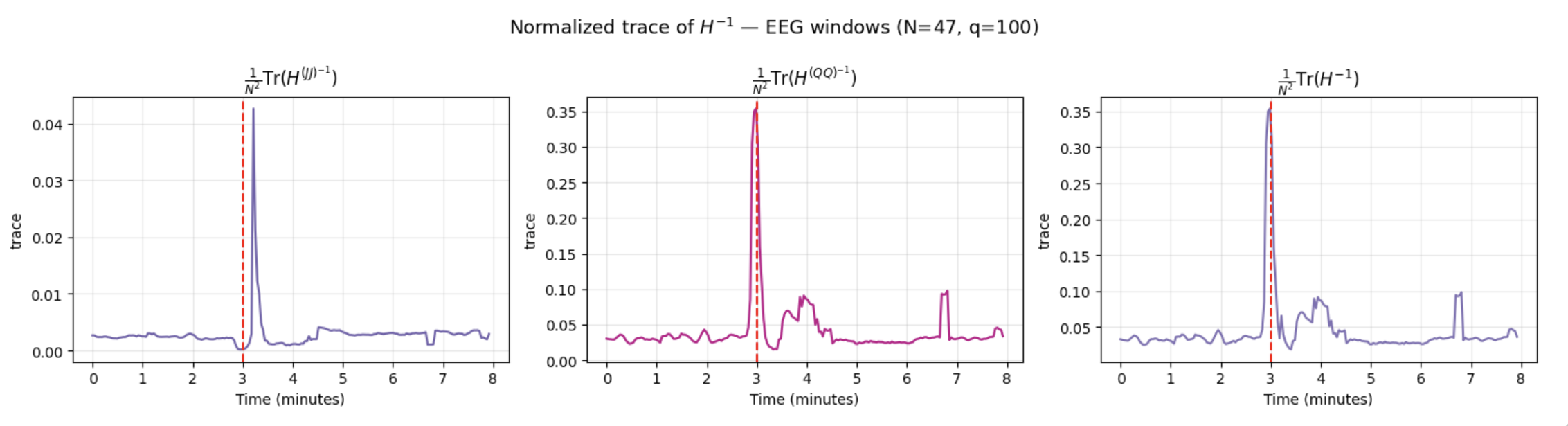}
    \caption{Normalized inverse-Hessian traces
    $\tfrac{1}{N^2}\Tr(\mathbf H^{(JJ)^{-1}})$ (left),
    $\tfrac{1}{N^2}\Tr(\mathbf H^{(QQ)^{-1}})$ (center), total (right) from
    EEG. The red dashed line marks seizure onset ($t\approx3$~min); the
    pronounced peak signals critical slowing down and the accompanying loss
    of statistical identifiability. $N=47$, $q=80$.}
    \label{fig:eeg_trace}
\end{figure}

\begin{figure}
    \centering
    \begin{minipage}{0.48\textwidth}
        \centering
        \includegraphics[width=\textwidth]{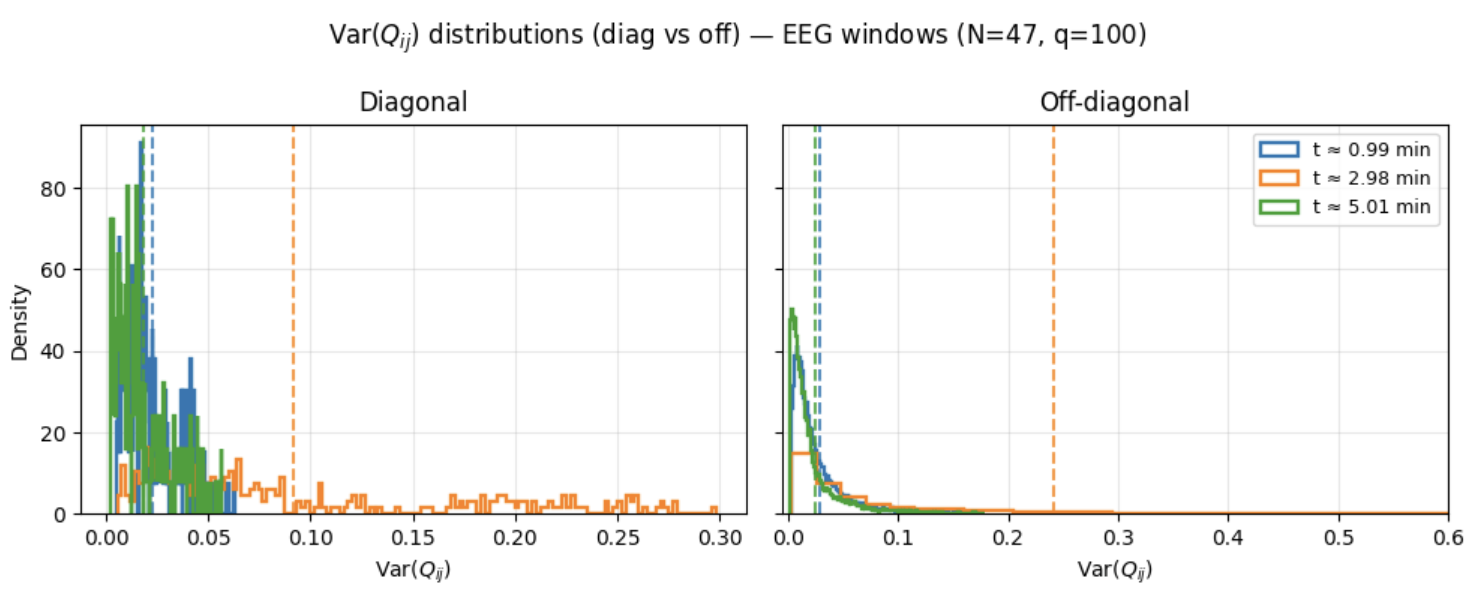}\\
        \small{(a) EEG, $N=47$, $q=100$}
    \end{minipage}\hfill
    \begin{minipage}{0.48\textwidth}
        \centering
        \includegraphics[width=\textwidth]{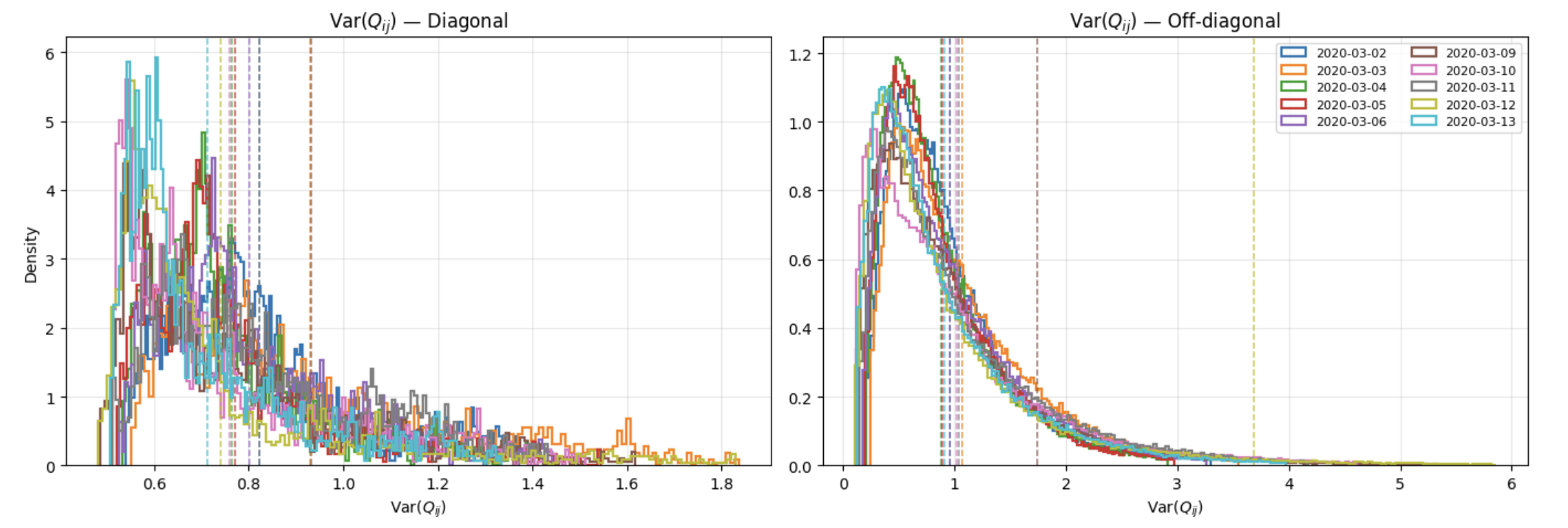}\\
        \small{(b) Finance, $N=47$, $q=80$}
    \end{minipage}
    \\[6pt]
    \begin{minipage}{0.48\textwidth}
        \centering
        \includegraphics[width=\textwidth]{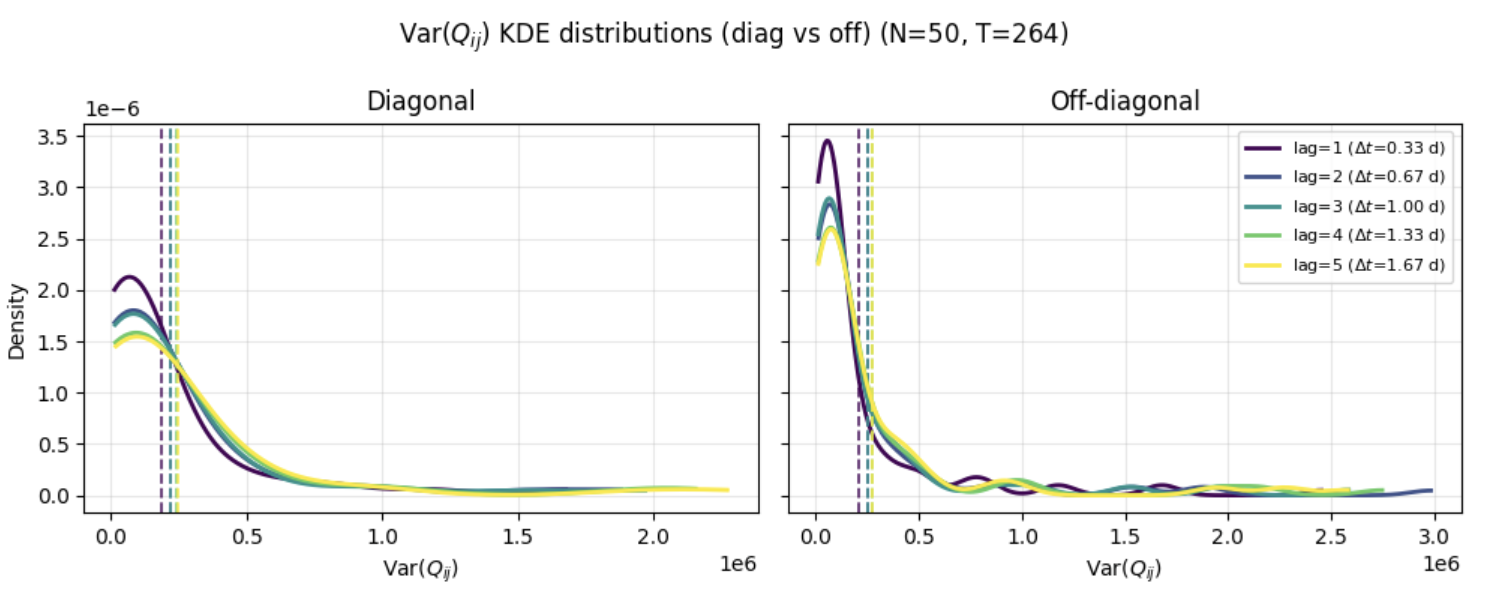}\\
        \small{(c) Plankton, $N=47$, $T=264$}
    \end{minipage}
    \caption{Distributions of $\mathrm{Var}(Q_{ij})\simeq(\mathbf H^{-1}_{QQ})_{ijij}$
    for diagonal (left within each panel) and off-diagonal (right) couplings.
    (a) EEG, by temporal stage (preictal, onset, ictal). (b) Finance, by
    trading day. (c) Plankton, by sampling lag $\Delta t\in[0.33,1.67]$~days;
    the inferred $\hat c\simeq0.988$ places the ecosystem close to
    criticality.}
    \label{fig:varQ_data}
\end{figure}

\end{document}